\documentclass[]{article}
\usepackage{cite}
\usepackage{amsmath,amssymb,amsfonts}
\usepackage{algorithmic}
\usepackage{graphicx}
\usepackage{textcomp}

\usepackage{subfig}
\DeclareGraphicsExtensions{.pdf,.jpeg,.png}
\usepackage{url}
\usepackage{multirow}
\usepackage{color}
\usepackage{array}

\usepackage{booktabs}
\usepackage{threeparttable}

\providecommand{\zxhrefeq}[1]{(\ref{#1})}

\providecommand{\Zxhreftb}[1]{Table~\ref{#1}}

\newcommand{\tabincell}[2]{\begin{tabular}{@{}#1@{}}#2\end{tabular}}
 
\definecolor{color1}{rgb}{0.8,0.1,0.9}
\definecolor{color2}{rgb}{0.9,0.4,0.1}
\definecolor{color3}{rgb}{0.1,0.6,0.15}

\def\BibTeX{{\rm B\kern-.05em{\sc i\kern-.025em b}\kern-.08em
    T\kern-.1667em\lower.7ex\hbox{E}\kern-.125emX}}
\begin{document}
\title{Unsupervised Domain Adaptation with Variational Approximation for Cardiac Segmentation}
\author{Fuping~Wu and Xiahai~Zhuang 
\thanks{This work was funded by the National Natural Science Foundation of China (grant no. 61971142, 11871165, and 62111530195). Xiahai~Zhuang is the corresponding author.}
\thanks{Fuping~Wu is a PhD student with the Department
	of Statistics, School of Management, and School of Data Science, Fudan University, 200433, Shanghai, China.}
\thanks{Xiahai~Zhuang is with School of Data Science, Fudan University (e-mail: zxh@fudan.edu.cn).}}

\maketitle

\begin{abstract}
Unsupervised domain adaptation is useful in medical image segmentation. Particularly, when ground truths of the target images are not available, domain adaptation can train a target-specific model by utilizing the existing labeled images from other modalities. 
Most of the reported works mapped images of both the source and target domains into a common latent feature space, and then reduced their discrepancy either implicitly with adversarial training or explicitly by directly minimizing a discrepancy metric.
In this work, we propose a new framework, where the latent features of both domains are driven towards a common and parameterized variational form, whose conditional distribution given the image is Gaussian. This is achieved by two networks based on variational auto-encoders (VAEs) and a regularization for this variational approximation.
Both of the VAEs, each for one domain, contain a segmentation module, where the source segmentation is trained in a supervised manner, while the target one is trained unsupervisedly.
We validated the proposed domain adaptation method using two cardiac segmentation tasks, i.e., the cross-modality (CT and MR) whole heart segmentation and the cross-sequence cardiac MR segmentation.
Results show that the proposed method achieved better accuracies compared to two state-of-the-art approaches and
demonstrated good potential for cardiac segmentation.
Furthermore, the proposed explicit regularization was shown to be effective and efficient in narrowing down the distribution gap between domains, which is useful for unsupervised domain adaptation.
Our code and data has been released via \url{https://zmiclab.github.io/projects.html}.

\end{abstract}

\section{Introduction}
Accurate cardiac segmentation is an essential prerequisite for many medical applications, such as 3D modeling and  functional analysis, which are important for diagnosis of cardiac diseases \cite{KARAMITSOS20091407,petitjean2011review}. In clinics, multi-modality medical images are widely used to assist diagnosis.
However, obtaining the automated segmentation for all modality images can be label intensive and expensive.
To alleviate this, an effective learning-based approach is to train an automatic segmentation model using existing labeled data of one modality and adapt this model for the automatic segmentation of the other modalities.
However, the performance of such adapted segmentation models usually degrades significantly when they are tested on images from new modalities, due to \textit{dataset bias}, which is also known as \textit{domain shift} \cite{shimodaira2000improving}.

\begin{figure*}[!t]
	\centering
	\subfloat[The framework of previous researches of domain adaptation in a latent feature space. The domains were mapped into a common latent feature variable $z$. The domain discrepancy was then either reduced by adversarial training or explicitly minimization of discrepancy metrics.]{\includegraphics[width=3.3in]{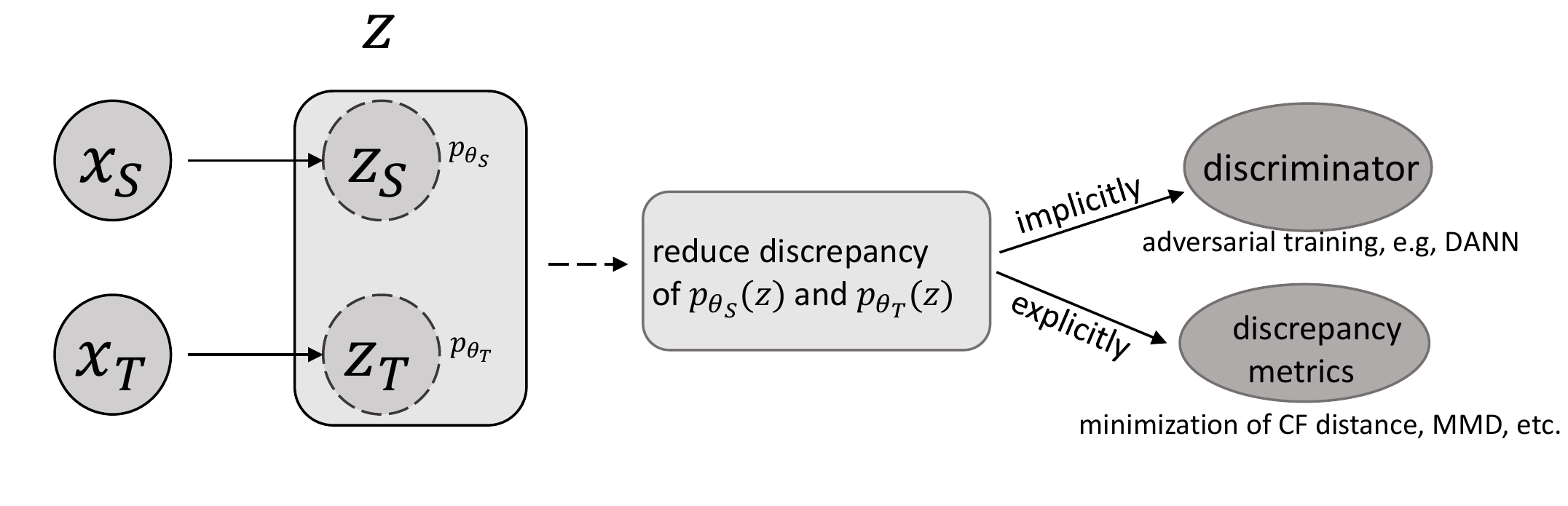}
		\label{fig:previous idea}}
	\hspace{0.055in}
	\subfloat[The idea of the proposed domain adaptation method via variational approximation. The two domains were driven towards parameterized distributions $q_{\phi_{S}}$ and $q_{\phi_{S}}$ via variational auto-encoder. A regularization term was proposed to constrain $q_{\phi_{S}}$ and $q_{\phi_{S}}$ to be the same distribution. ]{\includegraphics[width=3.3in]{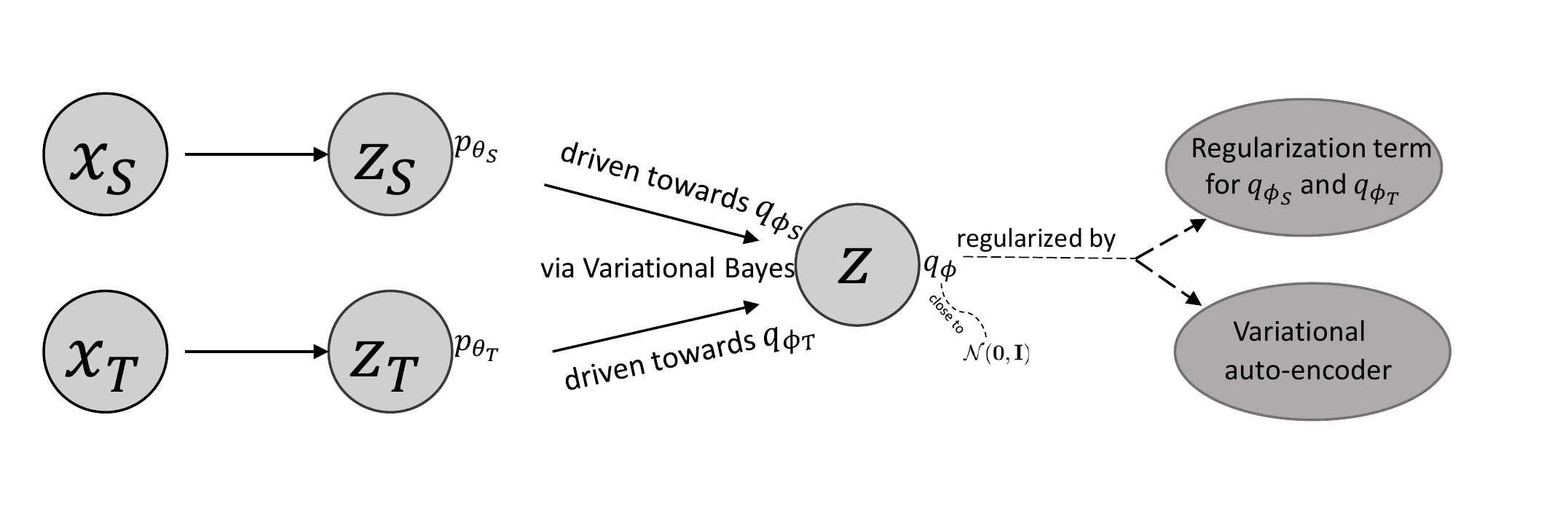}
		\label{fig:this work idea}}
	\caption{Illustration of the difference between the previous works and the proposed method for domain adaptation in a latent feature space. $x_S$ and $x_T$ denote the input source and target image, respectively. $z$ is the latent feature variable. $p_{\theta_S}$ and $p_{\theta_T}$ are probability functions on the source and target data, parameterized with $\theta_S$ and $\theta_T$, respectively. $q_\phi$ is the variational approximation of $p_{\theta_{S/T}}$.}
	\label{fig:difference}
\end{figure*}

A common approach is to fine-tune the  models using labeled images from the new modalities \cite{babenko2014neural,chu2016best,oquab2014learning}, which however requires extra manually annotated training data. 
Recently, researchers have proposed a number of new methodologies \cite{raina2007self,thrun2012learning}, among which domain adaptation, a special technique of transfer learning, has became increasingly popular, thanks to its needlessness of additional labeled images from the target modalities \cite{Csurka2017}.

Domain adaptation transforms the images from different modalities into a modality-invariant common space \cite{dou2018pnp}, or directly translates the images from one modality to another with the same anatomic structures \cite{chen2019synergistic}.
Formally, the imaging modality with ground truth segmentation is referred to as the \textit{source domain}, and the modality without ground truth, which is our target for automated segmentation, is denoted as the \textit{target domain}.
Correspondingly, the data from these two domains are known as the source data and target data, respectively.
In this work, we follow the idea of aligning the distributions of these domains in a latent feature space,
which is equivalent to extract modality-invariant features in this specific task  \cite{fernando2013unsupervised,sun2016return,tommasi2013frustratingly}.
Hence, domain adaptation becomes a problem of feature learning and minimization of domain discrepancy.

The latent features can be learned adaptively by deep neural networks (DNNs).
Particularly, the generative adversarial networks (GANs) have demonstrated great potential \cite{arjovsky2017wasserstein,ganin2016domain,goodfellow2014generative}.
By introducing discriminators, the discrepancy between domains can be minimized implicitly.
This adversarial training has also been  adopted in medical image analysis \cite{kamnitsas2017unsupervised,dou2018pnp,chen2019synergistic}.
However, this technique is still challenging, due to the difficulty of obtaining the nash equilibrium point in GANs when they are applied to domain adaptation \cite{heusel2017gans}.
Also, training the generator and discriminator networks for the implicit minimization of domain discrepancy could be complex \cite{arjovsky2017wasserstein}.
For cardiac segmentation, PnP-AdaNet \cite{dou2018pnp} and SIFA \cite{chen2019synergistic} are two most related works. PnP-AdaNet utilized features from multi-layer for adaptation, and SIFA minimized the domain discrepancy on both feature and image levels. They were both validated to be effective on cardiac dataset.

There were a few works reported adopting deep learning to developing explicit measurements for domain discrepancy \cite{long2015learning,long2017deep,sun2016deep,tzeng2014deep}, which were implemented by introducing moment statistics or unbiased estimators.
These works obtained promising results for classification tasks, however the application to segmentation has not been fully explored.
Recently, an explicit distance metric based on characteristic functions of distribution was proposed for domain adaptation \cite{Wu2020TMI}. This work, different from the adversarial training scheme, minimized the explicit metric directly to reduce discrepancy between the latent variables from the two domains.

		In this work, we propose a new framework of domain adaptation for cardiac image segmentation via the strategy of variational approximation (VarDA).
		The idea is illustrated in Figure \ref{fig:this work idea}, which shows a different solution from previous works.
		Specifically, instead of reducing the discrepancy of the latent features from two domains directly, we force the distributions of these features approximate to a parameterized probability distribution function.
		The approximation process can be achieved by  a variant of variational auto-encoder (VAE).
		Then, the distance between the two approximations can be taken as a regularization, which is validated to be effective for domain adaptation in experiments.

The VAE is adapted to work for both the labeled source data and unlabeled target data.
For segmentation, the VAE incorporates a prediction model, which takes only latent features as input, and thus can be shared by both domains.
We further deduce a new form of the variational lower bound for the log-likelihood of the labeled data.
This lower bound can be taken as an objective function to train the VAE models, and thus is used to form the total loss for the domain adaptation and segmentation task.

The main contributions of this work are summarized as follows:
(1) We develop a new framework for domain adaptation, where the latent features from both domains are approximated by
		a common and parameterized variational form. The framework is implemented using deep neural networks based on variational auto-encoder (VAE).		
		(2) We propose an explicit regularization, which is computed from the distance between the approximations for the distributions of latent variables.
(3) We validate our proposals using experiments involving two cardiac image segmentation tasks.
In the experiments, comparisons with other state-of-the-art algorithms are provided, and parameter studies are performed.


\section{Related work}\label{related}
This work is mainly related to two active research areas, i.e., domain adaptation and variational autoencoder (VAE).

\subsection{Domain adaptation}

For domain adaptation, a particular scenario is when we have the paired data, namely each image from one modality (domain) has the corresponding image of the same subject from the other modality (domain).
Chartsias et al. \cite{chartsias2017multimodal} studied a multi-input multi-output fully convolutional neural network for MR synthesis. They combined latent features from all modalities into a single fused representation, which can be transformed into any target modality with a decoder.
Tulder et al. \cite{van2019learning} proposed a shared autoencoder-like convolutional network, which learns a common
representation from multi-modal data.
They utilized feature normalization and modality dropout techniques to improve the cross-modality classification accuracy.
The core idea in these two works was to force the latent features from paired data to be similar. To learn the models, they processed the training data pairwisely, which does not need to consider the difference of data distributions from modalities.
Moreover, Shi et al. \cite{conf/nips/ShiNPT19} proposed the mixture-of-experts of VAE, which implicitly decomposed multi-modal data into shared and private subspaces, and could be applied for complex heterogeneous modalities, such as images and the language data.
Antelmi et al. \cite{conf/icml/AntelmiARL19} adopted multi-channel VAE to tackle paired heterogeneous data, and extracted parsimonious and interpretable representations by variational dropout.

For unpaired data, extracting modality-invariant features is particularly challenging. As deep learning methods have been widely used in domain adaptation, here we mainly focus on DNN-based approaches.
Researchers have proposed explicit metrics to measure domain discrepancy.
Tzeng et al.\cite{tzeng2014deep} and Long et al.\cite{long2015learning} employed the Maximum Mean Discrepancy (MMD) as a domain confusion metric, and minimized the MMD loss together with a task-specific loss to learn the modality-invariant representations.
Sun et al.\cite{sun2016deep} proposed a new method, referred to as Deep CORAL in their paper, to align the second-order statistics of the source and target distributions.
Instead of using the Euclidean distance in Deep CORAL, Wang et al.\cite{wang2017deep} proposed to adopt the Riemannian distance of covariance matrices between the source and target domains, which can be approximated by the Log-Euclidean distance.
While these methods were solely developed for classification tasks, recently,  Wu and Zhuang\cite{Wu2020TMI} proposed the CFD method for medical image segmentation, which explicitly calculated the domain discrepancy with the distance between the characteristic functions of the distributions in a latent feature space.
		Although CFD was validated to be effective, it requires to be combined with other techniques to achieve a comparable performance to the adversarial training methods. These techniques include the mean value matching and image reconstruction. Moreover, it was only tested on simple segmentation tasks of two or three structures. While for the whole heart segmentation which has much more complex structures, CFD could be challenged to achieve successful unsupervised domain adaptation due to the complexity of the task.
Besides these statistic estimators, graph matching based metrics were also studied.
For example, Das and Lee \cite{Debasmit2018Graph} first constructed the representation graphs for both source and target domains, and then minimized the matching loss between them, which consisted of node-to-node matching and edge-to-edge matching.
Yang and Yuen \cite{conf/aaai/YangY19} constrained the edges in both graphs for two domains, and mapped the data into features with unified structure criteria via adversarial training.
Pilanci and Vural \cite{9055059Pilanci} matched the Fourier bases of their graph instead of the nodes, such that the label functions on the two graphs can be reconstructed with similar coefficients.

GAN has recently been shown to have great potential for domain adaptation, particularly for medical image generation and segmentation.
Zhang et al.\cite{zhang2018translating} utilized cycle-consistent generative adversarial networks (CycleGANs) to achieve cross-modality synthesis between MRI and CT images.
To reduce the geometric distortion, they introduced a shape-consistency loss for the generators.
Combining the idea of style transfer, Huang et al.\cite{huang2018multimodal} proposed to decompose image features from both domains into disentangled representation of domain-invariant content and domain-specific style, which were then used to simulate target images.
They employed GANs to align the distributions of these translated target images with the real ones, and the alignment was implemented on image levels.

For classification or segmentation tasks, domain adaptation was mainly conducted on feature levels.
Without disentangling features, data from different domains were mapped into a common latent code, which was then forced to be domain-invariant.
Kamnitsas et al.\cite{kamnitsas2017unsupervised} proposed a so-called multi-connected architecture for the domain discriminator, where the multi-layer features were connected before being input into the discriminator.
They obtained evident improvement in MRI-CT cross-modality segmentation of brain lesions.

Regarding to  segmentation  of domain adaptation, one of the most related works to ours is from Dou et al.\cite{dou2018pnp}. They proposed a plug-and-play domain adaptation module, based on a dilated fully convolutional network, to map two domains into a common space. Their framework was a modified version of the domain adversarial neural networks (DANN) \cite{ganin2016domain}. It was validated on  MRI-CT cross-modality segmentation of 2D cardiac images.
Another related state-of-the-art work is from Chen et al.\cite{chen2019synergistic}, whose method is referred to as SIFA.
The authors implemented adaptation at both of the image and feature levels via adversarial training, and used the cycle-consistency loss.
This work was reported to outperform peer methods for natural images by a significant margin.
More recently, Ouyang et al.\cite{ouyang2019data}  proposed a data efficient method for multi-domain medical image segmentation.
They combined a VAE-based feature prior matching and domain adversarial training to learn a
shared latent space which is expected to be domain-invariant.
These state-of-the-art methods designed for medical image segmentation are all based on adversarial training, which aligns the distributions of two domains by minimizing their discrepancy implicitly.

\subsection{Variational autoencoder}

VAE is a popular deep generative model. Using VAE, one can approximate the posterior distribution of a latent variable  conditioned on data using a normal distribution \cite{kingma2013auto}. This property is particularly useful, and enables us to drive two domains towards a common and parameterized variable in a latent space.
Kingma et al.\cite{kingma2014semi} developed a new model based on VAE for semi-supervised learning with small labeled data sets.
The model allowed for effective generalization to large unlabeled ones, which were assumed to be from the same distributions with the labeled data.
Furthermore, based on conditional variational autoencoder  with Gaussian latent variables,  Sohn et al.\cite{sohn2015learning} proposed a stochastic neural network to make inference for structured output variables.
They performed the deterministic inference by maximizing a generation network, which uses both the data and its corresponding latent features for prediction.
Another important VAE work was from Walker et al.\cite{walker2016uncertain}, where the authors constructed a VAE model to predict dense trajectories from pixels.
The method employs an encoder to estimate the posterior distributions of the latent variables conditioned on the data and labels, and a decoder to predict the conditional distributions of trajectories given images.

In this work, the distributions of the source data (with labels) and target data (without labels) are different, since the domain shift exists.
Also, the VAE model, such as the one proposed by Walker et al.\cite{walker2016uncertain}, generally requires labels for all data, which however are not available for the target data in the unsupervised domain adaptation task.
Therefore, in this work we propose a new form of VAE for domain adaptation, and deduced an explicit regularization  for domain discrepancy.

\begin{figure*}[t]
	\centering
	\includegraphics[width=0.78\textwidth]{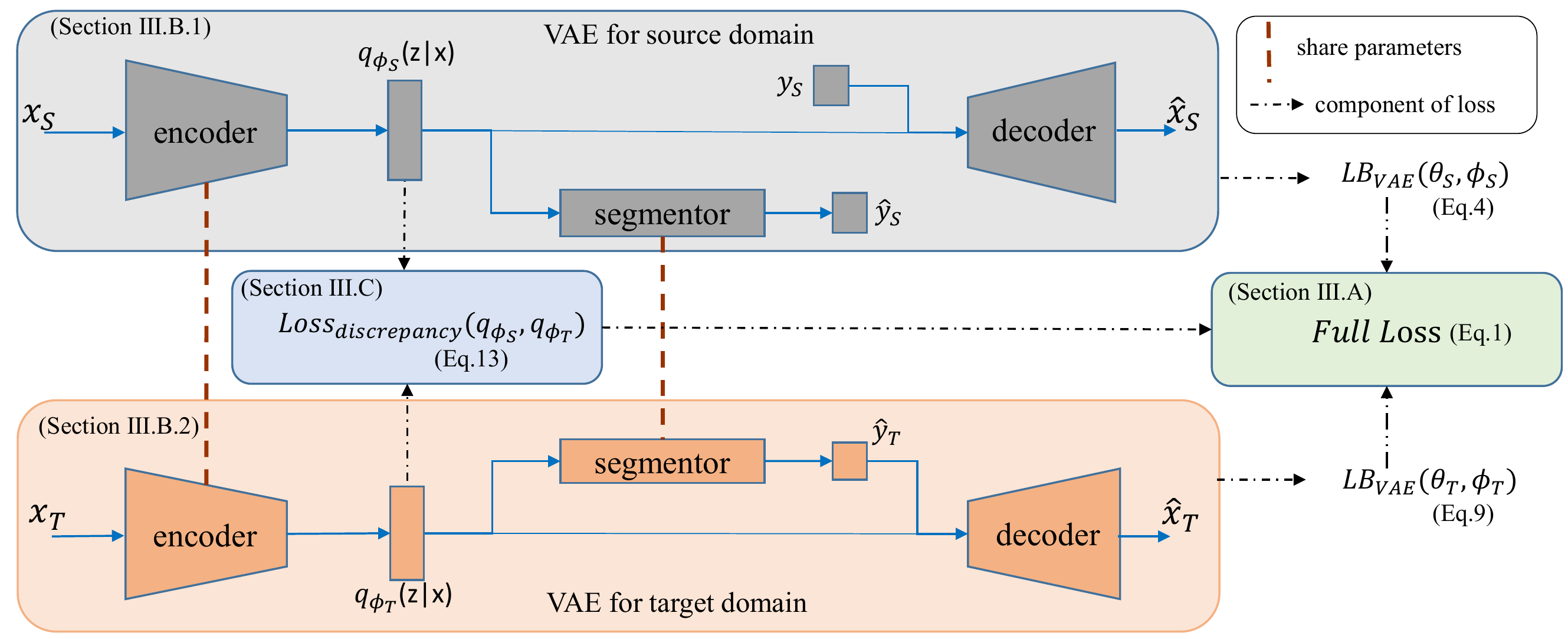}
	\caption{Framework of the proposed domain adaptation method. For each domain, we construct a modified VAE model, which contains an encoder to extract modality-invariant latent features, a decoder for image reconstruction, and a segmentor for image segmentation. The domain discrepancy is computed explicitly with the estimated distributions of the latent features from two domains.
	}
	\label{fig:framework}
\end{figure*}

\begin{table}
	\centering
	\caption{Reference to the mathematical symbols. }
	\label{method:notation}
	\begin{tabular}{|c|p{5cm} |}
		\hline
		subscript $_{S/T}$ & variables from source/ target domain  \\\hline
		$I_{S/T}$; $x$ & image domain; image variable \\\hline
		$y$, $z$& segmentation variable, latent variable \\\hline
		$N_{S/T}$& number of samples from source/ target domain\\ \hline
		$x_{S/T}^i,y_{S}^i,\widehat{y}_{T}^i$& image sample, ground truth segmentation, and predicted segmentation\\\hline
		$X_{S/T},Y_{S},\widehat{Y}_{T}$& the sets of the above samples \\\hline
		$p_{\theta}()$& PDF of model variables with parameter $\theta$  \\\hline
		$q_{\phi}()$&  approximation of $p_{\theta}$ with parameter $\phi$\\ \hline
		$D(\phi_S,\phi_T)$& distance metric of distributions\\	\hline
		$\mathcal{D}(\phi_S,\phi_T)$& distance estimated from marginal distributions\\ \hline
		$\widetilde{\mathcal{L}}()$& Monte Carlo estimate of VAE objective function\\ \hline
		$\widetilde{\mathcal{H}}()$& minibatch-based Monte Carlo estimate of total loss function\\\hline
	\end{tabular}
\end{table}

\section{Methodology} \label{method}

In this work, we aim to align the two domains in a common latent feature space.
We first transform each  domain into a latent feature variable.
		The two variables are then driven and approximated by a common and parameterized variational form via VAEs.
		As this approximation can be estimated with either the source or target data, we obtain two forms of estimation for the distribution of the common variational form.
		The distance of the two estimations is then used as an effective regularization for domain adaptation.
Then, we train the segmentation model with the latent feature and the ground truth of source data.
The resulting model is expected to be applicable to the segmentation of the target images.

We denote the independent and identically distributed ($i.i.d.$) samples from the source image domain $I_S$, as $X_S=\{x_S^i\}_{i=1}^{N_{S}}$, where $x_S^i$ is one of the $N_S$ samples.
The corresponding segmentation of $x_S^i$ is denoted as $y_S^i$,  which can be considered as $i.i.d.$ samples of a label variable, $y_S$.
Similarly, $X_T=\{x_T^i\}_{i=1}^{N_T}$ are $N_{T}$ $i.i.d.$  samples from the target image domain $I_T$,
whose segmentation, $Y_T=\{y_T^i\}_{i=1}^{N_T}$, is unknown.
Here, the symbols with subscript $S$ indicate the variables or samples are for the source domain, and those with $T$ are for the target domain.
We denote the probability density functions (PDFs) of the latent variable $z_S$ and $z_T$ as $p_{\theta_S}(z)$ and $p_{\theta_T}(z)$, for which $\theta_S$ and $\theta_T$ are model parameters to be learned.
The posterior of $z_S$ w.r.t. the source image variable $x_S$ is   $p_{\theta_S}(z_S|x_S)$.
Hereafter, when a single symbol is enough to indicate which domain the variables or samples belong to, we omit the others. For example, we denote $p_{\theta_S}(z_S|x_S)$ as $p_{\theta_S}(z|x)$ for simplicity.
\Zxhreftb{method:notation} provides the reference of the key symbols used in this paper.

In the rest of this section, we first introduce the framework of the proposed domain adaptation method in Section \ref{framework},
where the total loss function is given.
Then, we elaborate on the VAE models, which are formulated to incorporate segmentation models, in Section \ref{modifiedvae}.
Here, we deduce a variational lower bound as the objective function.
	Finally, based on the modified VAE, we propose an explicit metric as a regularization for domain adaptation, in Section \ref{measurement}.

\subsection{The proposed framework for domain adaptation} \label{framework}


Fig.\ref{fig:framework} illustrates the proposed framework, which consists of three components,
i.e., (1) the VAE for source domain, (2) the VAE for target domain, and (3) the module for regularization on domain discrepancy.

The source domain VAE is composed of an \textit{encoder} for feature extraction, a \textit{decoder} for image reconstruction, and a \textit{segmentor} for segmentation.
The encoder maps the image $x_S$ into a shared latent feature space, and approximates the posterior probability $p_{\theta_S}(z|x)$ by a parameterized model $q_{\phi_S}(z|x)$.
The random variables in the shared latent space are denoted as $z_S$ and $z_T$ for the source and target domains, respectively.
At the training stage, we sample the latent features from the posterior $q_{\phi_S}(z|x)$, and feed them into the networks of segmentor and decoder.
The segmentor outputs the probability of label category
for each pixel,
and the decoder, conditioned on the segmentation label, reconstructs the input image.

The target domain VAE has a similar structure to the source domain, except that it uses the predicted segmentation  as the label input of the decoder.
We denote the objective functions of the two VAEs as $LB_{VAE}(\theta_S,\phi_S)$ and $LB_{VAE}(\theta_T,\phi_T)$, respectively. The detail is given in Section \ref{modifiedvae}.

For the domain discrepancy, one can estimate the two variational approximations for  $z_S$ and $z_T$, i.e., $q_{\phi_S}(z)$ and $q_{\phi_T}(z)$, from which the input data of the segmentor are sampled, based on the two VAE modules mentioned before.
To force the two approximations to be the same distribution, we therefore propose a regularization for this discrepancy by computing the distance between $q_{\phi_S}(z)$ and $q_{\phi_T}(z)$.
The loss function is denoted as $Loss_{discrepancy}(q_{\phi_S},q_{\phi_T})$ and will be discussed in Section \ref{measurement}.

With the aforementioned three components, the total loss function of the domain adaptation segmentation method is then formulated by,
\begin{equation}\label{2.1}\begin{array}{l@{\ }l}
Full\,Loss(\omega)=&-\alpha_1 LB_{VAE}(\theta_S,\phi_S) -\alpha_2 LB_{VAE}(\theta_T,\phi_T)\\
&+\alpha_3 Loss_{discrepancy}(q_{\phi_S},q_{\phi_T}),
\end{array}\end{equation}
where $\omega=(\theta_S,\phi_S,\theta_T,\phi_T)$ are the parameters to be optimized, and $\alpha_1$, $\alpha_2$, $\alpha_3$ are the parameters controlling the trade-off between the three terms.

At the test stage, we input a test image $x_T$ into the encoder, and obtain a distribution $q_{\phi_T}(z|x)$. We calculate the mean value of this distribution as the input of the trained segmentor, which outputs the predicted segmentation $\widehat{y}_T$.

\subsection{Conditional variational autoencoder with segmentation}
\label{modifiedvae}
In the proposed framework, the VAE models need to incorporate the function of segmentation, as our task is cardiac image segmentation.
This is in contrast to the widely used VAE in \cite{kingma2013auto}, where the decoding is only for the reconstruction of images.
Therefore, we propose to maximize the joint log-likelihood (JLL) of the complete data as the objective function of VAE, i.e., $\mathrm{JLL}=\log p_{\theta} (X,Y) $. The complete data consist of both the images and their segmentation labels.
This is  the major difference to the VAE in \cite{kingma2013auto}, whose objective function is based on the log-likelihood of the images only.

Furthermore, since we do not have the gold standard segmentation of the target images,  a new formulation is needed to tackle the missing labels for the VAE of target domain.
We introduce this formulation in  Section~\ref{method:vaetarget}, where the corresponding objective function is deduced.



\subsubsection{VAE for source domain}\label{method:vaesource}

The proposed VAE maximizes the joint log-likelihood of the complete data. For the source data, we have
\begin{displaymath}
\mathrm{JLL}=\log p_{\theta_S}\big((x_S^1,y_S^1), \cdots,(x_S^{N_S},y_S^{N_S})\big) .\end{displaymath}
As all data points can be taken as samples from $i.i.d.$ random variables, the joint log-likelihood becomes the sum over that of individual data points, i.e.,
\begin{align}
\mathrm{JLL}=\sum_{i=1}^{N_S}\log (p_{\theta_S}(x_S^i) p_{\theta_S}(y_S^i|x_S^i)). 
\label{2.6}		
\end{align}

To optimize this generative model $p_{\theta_S}(x,y)$, a commonly adopted method is to introduce a latent variable $z$.
As $p_{\theta_S}(z|x)$ is generally intractable, we employ VAE to approximate it by a parameterized model $q_{\phi_S}(z|x)$, which can be implemented using neural networks easily.
To infer the discriminative model $p_{\theta_S}(y|z)$, we follow the assumption of distribution independence  in \cite{kingma2014semi}, i.e., $q_{\phi_S}(y,z|x)= q_{\phi_S}(y|x)\cdot q_{\phi_S}(z|x)$.

Under this assumption, we can obtain a new form of variational lower bound for $\log p_{\theta_S}(x,y)$, denoted as $LB_{VAE}(\theta_S,\phi_S)$, i.e.,
\begin{align}
\log p_{\theta_S}(x,y) &\ge LB_{VAE}(\theta_S,\phi_S) .
\label{2.7}
\end{align}
Here, $LB_{VAE}(\theta_S,\phi_S)$ can be formulated by,
\begin{align}
&LB_{VAE}(\theta_S,\phi_S)=-D_{KL}(q_{\phi_S}(z|x)\parallel p_{\theta_S}(z))\quad \notag \\
& +E_{\log q_{\phi_S}(z|x)}[ p_{\theta_S}(x|y,z)]+ E_{q_{\phi_S}(z|x)}[\log p_{\theta_S}(y|z)],
\label{2.8}
\end{align}
where $D_{KL}(q || p)$ is the $KL$ divergence of $q$ and $p$.
The second term $E_{q_{\phi_S}(z|x)}[\log p_{\theta_S}(x|y,z)]$ indicates the log-probability of the input image $x$ given the label $y$ and feature $z$, which is conditioned on $x$. This term can be modeled by the image reconstruction process.
		The third term $E_{q_{\phi_S}(z|x)}[\log p_{\theta_S}(y|z)]$ indicates the log-probability of the segmentation prediction $y$ given $z$, which is conditioned on $x$. This term is modeled by the segmentor in the proposed framework.
The proof for this lower bound can be found in the Supplementary Material. 
By reparameterization using a differentiable transformation $g_{\phi_S}(\epsilon,x)$ of an (auxiliary) noise variable $\epsilon$  \cite{kingma2013auto}, i.e.,
\begin{equation}
z=g_{\phi_S}(\epsilon,x)  ~~~~~~~   with ~~ \epsilon\sim p(\epsilon),
\label{2.9}
\end{equation}
we form a Monte Carlo estimate of $LB_{VAE}(\theta_S,\phi_S)$ in \zxhrefeq{2.8}, i.e.,
\begin{align}
\widetilde{\mathcal{L}}_S(\theta_S,&\phi_S;x^i,y^i)=-D_{KL}(q_{\phi_S}(z^i|x^i)\parallel p_{\theta_S}(z^i))\notag \\
&+\frac{1}{L}\sum_{l=1}^{L}\big[\log p_{\theta_S}(x^i|y^i,z^{(i,l)})+\log p_{\theta_S}(y^i|z^{(i,l)})\big],
\label{2.12}
\end{align}
where $z^{(i,l)}=g_{\phi_S}(\epsilon^{(i,l)},x^i)$, with $\epsilon^{(i,l)}\sim p(\epsilon)$, and $L$ is the number of samples.

For Eq.\zxhrefeq{2.8}, $D_{KL}(q_{\phi_S}(z|x)\!\!\parallel\!\!p_{\theta_S}(z))$ usually can be integrated analytically, and only the second and third right hand side terms require estimation by sampling. Specifically, let the prior over the latent variable $z_S$ be  multivariate Gaussian, i.e., $p_{\theta_S}(z)= N(0,\textbf{I})$, where $\textbf{I} \in R^{n\times n}$ is the identity matrix, and $q_{\phi_S}(z|x)=N(u_S(x),\Sigma_S(x))$, $u_S(x)=(u_{1}(x),\cdots,u_{n}(x))$, $\Sigma_S(x)={\rm diag}(\sigma_S(x))$,$\sigma_S(x)=(\lambda_{1}(x),\cdots,\lambda_{n}(x))$. For simplicity we denote $u_S(x), \Sigma_S(x)$ as $u_S, \Sigma_S$, then we have
\begin{align}
D_{KL}(q_{\phi_S}(z|x)\!\parallel\!p_{\theta_S}(z))=\frac{1}{2}\sum_{j=1}^{n}(\lambda_{j}+u_{j}^{2}-\log \lambda_{j}-1).
\end{align}

For complete source dataset, we have the estimator of the lower bound, based on mini-batches of $M$ data points, i.e.,
\begin{equation}
\widetilde{\mathcal{L}}_S(\theta_S,\phi_S;X_S,Y_S)=\frac{N_S}{M}\sum_{i=1}^{M}\widetilde{\mathcal{L}}_S(\theta_S,\phi_S;x^{i},y^{i}) .
\label{2.14}
\end{equation}

As Fig. \ref{fig:framework} illustrates, the module of \textit{VAE for source domain} consists of three parts, i.e.,
the encoder for approximation $q_{\phi_S}(z|x)$,
the decoder modeling the reconstruction term $E_{q_{\phi_S}(z|x)}[\log p_{\theta_S}(x|y,z)]$,
and the segmentor learning from the prediction term $E_{q_{\phi_S}(z|x)}[\log p_{\theta_S}(y|z)]$ in Eq.\zxhrefeq{2.8}.
As the reconstruction term indicates the reconstruction accuracy, we minimize the Mean-Squared Loss (MSELoss) between the input image and the reconstructed image from the decoder.
The prediction term indicates the prediction accuracy, thus we minimize  the MSELoss or cross entropy loss (CELoss) between the ground truth segmentation and the prediction from the segmentor.

\subsubsection{VAE for target domain}\label{method:vaetarget}

For the target data, we have a  lower bound similar to Eq.\zxhrefeq{2.8}, denoted as $LB_{VAE}(\theta_T,\phi_T)$,
where we however do not have ground truth segmentation.
We deal with this missing data using the pseudo labels, i.e., the predicted segmentation $\widehat{y}_T$ of $x_T$. The variational lower bound for the target domain becomes,
\begin{align}
&LB_{VAE}(\theta_T,\phi_T)=-D_{KL}(q_{\phi_T}(z|x)\parallel p_{\theta_T}(z))\notag \\
& +E_{q_{\phi_T}(z|x)}[\log p_{\theta_T}(x|\widehat{y},z)]+ E_{q_{\phi_T}(z|x)}[\log p_{\theta_T}(\widehat{y}|z)].
\label{2.15}
\end{align}

The  VAE structure for the target domain is presented in Figure \ref{fig:framework}. The segmentor shares parameters with that of the VAE for source domain. The decoder takes the latent features and prediction from the segmentor as input, and outputs the reconstructed images.

For the computation of Eq.\zxhrefeq{2.15}, we can obtain the Monte Carlo estimate of $LB_{VAE}(\theta_T,\phi_T)$ from the complete target data, similar to Eq.\zxhrefeq{2.12} and Eq.\zxhrefeq{2.14} for the source domain.
We denote this computation as $\widetilde{\mathcal{L}}_T(\theta_T,\phi_T;X_T,\widehat{Y}_T)$, where $\widehat{Y}_T=\{\widehat{y}_T^i\}_{i=1}^{N_T}$.
Also, we minimize the MSELoss between the input image and the reconstructed image from the decoder, to maximize the reconstruction term in Eq.\zxhrefeq{2.15}.

For the prediction term $E_{q_{\phi_T}(z|x)}[\log p_{\theta_T}(\widehat{y}|z)]$, one can use the MSELoss or CELoss. 
Here, MSELoss of the prediction itself equals zero,
and CELoss is equivalent to the conditional entropy loss \cite{grandvalet2005semi,shu2018a}.

\begin{table*}[!t]
	\centering
	\label{algorithm1}
	\begin{tabular}{l}
		\hline
		\textbf{Algorithm 1}. Optimizing $\widetilde{\mathcal{H}}$ based on mini-batches, and 
		computing $D(\phi_S,\phi_T)$ using Eq.\eqref{2.3}.\\
		\hline
		$\omega=(\theta_S,\theta_T,\phi_S,\phi_T)$ $\leftarrow$ Initialize parameters;\\
		Repeat\\
		~~~    1. $\{(x_S^{n_i},y_S^{n_i})\}_{i=1}^{M}$,$\{x_T^{n_j}\}_{j=1}^{M}$$\leftarrow$ Random mini-batch of $M$
		 data points; in source and target data, respectively ;\\
		~~~    2. Compute $u_S^{n_i},\Sigma_S^{n_i},u_T^{n_j},\Sigma_T^{n_j}$;\\
		~~~    3. $\epsilon_S^{(n_i,l)}, \epsilon_T^{(n_i,l)},\ l=1,\cdots L$ $\leftarrow$Random samples from 
		distribution $p(\epsilon)=N(\textbf{0},\textbf{I}), \textbf{I} \in R^{n\times n}$ ;\\
		~~~    4. Compute $z_S^{(n_i,l)}=(\Sigma_S^{n_i})^{\frac{1}{2}}*\epsilon_S^{(n_i,l)}+u_S^{n_i}$, and 
		$z_T^{(n_j,l)}=(\Sigma_T^{n_j})^{\frac{1}{2}}*\epsilon_T^{(n_j,l)}+u_T^{n_j}$ ;\\
		~~~   5. Compute $-\widetilde{\mathcal{L}}_S(\theta_S,\phi_S;x^{n_i},y^{n_i})$, $-\widetilde{\mathcal{L}}_T(\theta_T,\phi_T;x^{n_j},\widehat{y}^{n_j})$ 
		and $D(\phi_S,\phi_T)$ using Eq.\eqref{2.12}, Eq.\eqref{2.15} and Eq.\eqref{2.3},   \\ ~~~~~~
		respectively.  Their summation is taken as the loss function in Eq.\eqref{2.16};\\
		~~~    6. Update parameters using optimizer, such as SGD or 
		Adam; Until convergence of parameters $\omega$.\\
		Return $\omega$.\\
		\hline
		Test phase: the target images are fed into the target VAE, which outputs the segmentation predictions.\\
		\hline
	\end{tabular}
\end{table*}

\subsection{A regularization for the variational approximation}
\label{measurement}

Based on the VAE designed above, we can obtain two estimations, $q_{\phi_S}(z)$ and $q_{\phi_T}(z)$, for the distribution of the variational approximation.
Due to the domain shift between the two domains, the extracted features  $z_S$ and $z_T$ are subjected to different distributions. Hence, the approximations $q_{\phi_S}(z)$ and $q_{\phi_T}(z)$ are not the same one.
		To constrain them to be the same distribution, we define a regularization, i.e., $Loss_{discrepancy}(q_{\phi_S},q_{\phi_T})$ in Eq.\eqref{2.1}, by calculating the distance between them.
There exist different metrics for this distribution distance, including the Wasserstein distance \cite{arjovsky2017wasserstein} and the $l_p$ norm.
As the commonly used Wasserstein distance as well as KL divergence between $q_{\phi_S}(z)$ and $q_{\phi_T}(z)$ can not be integrated analytically for $M>1$, we adopted $l_2$ norm as a suitable metric, i.e., $\int[q_{\phi_S}(z)-q_{\phi_T}(z)]^2{\rm d}z$.
		Hereafter, we denote this distance metric as $D(q_{\phi_S}(z),q_{\phi_T}(z))$ or $D(\phi_S,\phi_T)$.
		Note that for the MMD method, the sampling operation for the latent features could lead to large variation of MMD estimator, and thus lead to ineffectiveness of the metric for discrepancy minimization \cite{Wu2020TMI}.

The distributions of the latent feature can be estimated via the modified VAE.
For the source data, as the VAE model  can  approximate the posterior probability $p_{\theta_S}(z|x)$ by a parameterized model $q_{\phi_S}(z|x)$ \cite{kingma2013auto,rezende2014stochastic},  the distribution of the approximation $q_{\phi_S}(z)$ can be estimated by,
\begin{align}
q_{\phi_S}(z)&\approx E_{p_{\theta_S}(x)}[q_{\phi_S}(z|x)].  \label{2.2}
\end{align}
Similarly, we introduce another VAE for the target data, and $q_{\theta_T}(z)$ can be estimated in the same way.

In this work, we adopt the minibatch strategy to optimize \zxhrefeq{2.1}. We randomly sample $M$ data points independently from the source domain and $M$ samples from the target domain, then $D(\phi_S,\phi_T)$ can be approximately calculated by,
\begin{flalign}
D&(\phi_S,\phi_T)=\int[q_{\phi_S}(z)-q_{\phi_T}(z)]^2{\rm d}z  \notag \\ 
&\approx \int\Big[\frac{1}{M}\sum_{i=1}^{M}q_{\phi_S}(z|x_S^i)-\frac{1}{M}\sum_{j=1}^{M}q_{\phi_T}(z|x_T^i)\Big]^2{\rm d}z \notag\\
&=\frac{1}{M^2}\sum_{i=1}^{M}\sum_{j=1}^{M}\left[k(x_S^i,x_S^j)+k(x_T^i,x_T^j)-2k(x_S^i,x_T^j)\right],
\label{2.3}
\end{flalign}
where $k(x_S^i,x_T^j)=\int q_{\phi_S}(z|x_S^i)\cdot q_{\phi_T}(z|x_T^j){\rm d}z$.

Let $N(u_S^i,\Sigma_S^i)$ or $N(u_T^j,\Sigma_T^j)$ be a normal distribution, which is adopted to model the posterior of the latent variables given the sample $x_S^i$ or $x_T^j$.
Particularly, we further adopt the simple situation, where $\Sigma_S^i={\rm diag}(\lambda_{S_1}^i,\cdots,\lambda_{S_n}^i)$, $\Sigma_T^j={\rm diag}(\lambda_{T_1}^j,\cdots,\lambda_{T_n}^j)$, $\lambda_{S_l}^i>0$ and $\lambda_{T_l}^j>0$, one can obtain that
\begin{equation}
k(x_S^i,x_T^j)=\frac{e^{-\frac{1}{2}\sum_{l=1}^{n}\frac{(u_{S_l}^i-u_{T_l}^j)^2}{\lambda_{S_l}^i+\lambda_{T_l}^j}}}{(2\pi)^{\frac{n}{2}}\cdot (\prod_{l=1}^{n}(\lambda_{S_l}^i+\lambda_{T_l}^j))^{\frac{1}{2}}},
\label{2.5}
\end{equation}
where $u_{S_l}^i$ is the $l$-th element of $u_{S}^i$.

%
%

The computation of Eq.\zxhrefeq{2.5} involves consecutive multiplication operations, which may be problematic when using deep learning frameworks (such as TensorFlow, PyTorch) for optimization. When the dimension of $z$ is large, the computation error of Eq.\zxhrefeq{2.5} could be non-negligible, especially at the back propagation stage.
To avoid this problem, we propose to compute the distance of the marginal distributions of $z_S\sim q_{\phi_S}(z)$ and $z_T\sim q_{\phi_T}(z)$ instead, i.e.,
\begin{align}
\widetilde{\mathcal{D}}(\phi_S,\phi_T)=\sum_{i=1}^{n}
\int[q_{\phi_S}(z_i)-q_{\phi_T}(z_i)]^2{\rm d}z,
\label{eq2.55}
\end{align}
where $\widetilde{\mathcal{D}}$ indicates the sum of distance of marginal distributions, and $z_i$ is the $i$-th element of $z$, and $n$ is the dimension of latent variable $z$..
Note that minimizing the distance of two PDFs, i.e., $D(\phi_S,\phi_T)$ of Eq.\zxhrefeq{2.3}, is not equivalent to minimizing the distance of their marginal distributions, i.e., $\widetilde{\mathcal{D}}(\phi_S,\phi_T)$ in Eq.\zxhrefeq{eq2.55}. Here, we propose the latter to be a surrogate solution of the former.
As discussed in the work of \cite{Wu2020TMI}, this substitution could lead to an ineffective constraint.
		However, as the features adapted in VarDA are extracted via variational approximation to $\mathcal{N}(\mathbf{0},\mathbf{I})$, the effect of this substitution can be alleviated with weak correlations among the element variables of the features.
		Specifically, as we assume the priors $p_{\theta_S}(z)$ and $p_{\theta_T}(z)$ to be both Gaussian, and their covariance matrices be diagonal. This assumption leads to the fact that $D(\theta_S,\theta_T)=0$ if and only if $\widetilde{\mathcal{D}}(\theta_S,\theta_T)=0$.
		As $q_{\phi_S}(z)$ and $q_{\phi_T}(z)$ approximate $p_{\theta_S}(z)$ and $p_{\theta_T}(z)$, respectively, this alternative is reasonable as a measurement of the discrepancy.
Moreover, $\widetilde{\mathcal{D}}(\phi_S,\phi_T)$ can be taken as a sliced version of $D(\phi_S,\phi_T)$. The detail of this sliced distance can be found in the Supplementary Material document.

Finally, the minibatch version of the total loss, i.e., Eq.\zxhrefeq{2.1}, denoted as $\widetilde{\mathcal{H}}(\omega)$, is given by,
\begin{align}
\widetilde{\mathcal{H}}(\omega)=
&-\alpha_1\cdot\widetilde{\mathcal{L}}_S(\theta_S,\phi_S;X_S,Y_S)\notag \\
&-\alpha_2 \cdot \widetilde{\mathcal{L}}_T(\theta_T,\phi_T;X_T,\widehat{Y}_T) +\alpha_3 \cdot \widetilde{\mathcal{D}}(\phi_S,\phi_T). \label{2.16}
\end{align}

Algorithm 1 provides the procedure of
minimizing $\widetilde{\mathcal{H}}$.

\section{Experiments and results} \label{exp}

\begin{table*}
	\centering
	\caption{Quantitative evaluation results of the segmentation of 3D MR images from MM-WHS dataset, where the CT images were used as the source domain. Note: N/A means that the ASSD value cannot be calculated due to no prediction for that cardiac structure. The best result in each column of the unsupervised methods is in boldface.
	}
	\label{ct to mr table}
	\resizebox{350pt}{36pt}{\!\!\!\!\!\!\!\!\!\!\!
		\begin{tabular}{|l|c@{\ \ }c@{\ \ }c@{\ \ }c@{\ \ }c|c@{\ \ }c@{\ \ }c@{\ \ }c@{\ \ }c|}
			\hline
			\multirow{2}{*}{methods} & \multicolumn{5}{c|}{Dice (\%)} & \multicolumn{5}{c|}{ASSD (mm)}  \\
			\cline{2-11}
			&MYO  & LA & LV &RA &  RV  & MYO & LA  & LV &RA & RV  \\
			\hline
			NoAdapt&0.122$\pm$0.124   &4.33$\pm$7.47  &0.980$\pm$1.06  &0.488$\pm$0.695  &20.2$\pm$17.7   &21.8$\pm$1.68  &N/A  & 47.2$\pm$25.4 &39.3$\pm$13.9   &17.6$\pm$10.6    \\
			\hline
			PnP-Ada &25.3$\pm$16.2  &46.3$\pm$21.5  &48.7$\pm$21.3 &51.2$\pm$12.4  &46.6$\pm$5.05   &11.3$\pm$2.8 &27.9$\pm$6.85  &\textbf{7.75$\pm$2.79}  &21.2$\pm$10.5  &17.0$\pm$2.30      \\
			\hline
			SIFA&31.5$\pm$12.1  &54.4$\pm$6.09  &\textbf{64.3$\pm$14.3}  &44.1$\pm$22.6  &26.7$\pm$20.0   &11.4$\pm$3.11  &14.7$\pm$9.22    &15.5$\pm$4.04   &16.2$\pm$9.15     &21.3$\pm$8.66    \\
			\hline
			VarDA& \textbf{39.9$\pm$16.2} &\textbf{54.8$\pm$6.67}  &60.9$\pm$16.2  &\textbf{65.5$\pm$4.34}  &\textbf{57.0$\pm$11.0}    &\textbf{5.95 $\pm$2.48} &\textbf{11.0$\pm$1.02}  &8.54$\pm$2.70 &\textbf{7.10$\pm$0.776}   &\textbf{9.02$\pm$2.79}  \\
			\hline
		\end{tabular}
	}
\end{table*}

\begin{table*}
	\centering
	\caption{Quantitative evaluation results of the segmentation of 2D MR images  from MM-WHS dataset, where the CT images as the source domain. Note: N/A means that the ASSD value cannot be calculated due to no prediction for that cardiac structure.}
	\label{ct to mr slice table}
	\resizebox{350pt}{36pt}{\!\!\!\!\!\!\!\!\!\!\!
		\begin{tabular}{|l|c@{\ \ }c@{\ \ }c@{\ \ }c@{\ \ }c|c@{\ \ }c@{\ \ }c@{\ \ }c@{\ \ }c|}
			\hline
			\multirow{2}{*}{methods} & \multicolumn{5}{c|}{Dice (\%)} & \multicolumn{5}{c|}{ASSD (mm)}  \\
			\cline{2-11}
			&MYO  & LA & LV &RA &  RV  & MYO & LA  & LV &RA & RV  \\
			\hline
			NoAdapt&0.0811$\pm$0.272   &3.08$\pm$11.63  &0.00$\pm$0.00  &0.742$\pm$2.44  &23.9$\pm$29.2   &N/A  &N/A  & N/A &N/A   &N/A    \\
			\hline
			PnP-Ada &32.7$\pm$23.8  &49.7$\pm$30.4  &48.4$\pm$28.2 &62.4$\pm$17.6  &44.2$\pm$18.8   &6.89$\pm$2.96 &22.6$\pm$11.1  &9.56$\pm$5.14    &20.7$\pm$16.0  &20.0$\pm$19.6       \\
			\hline
			SIFA&37.1$\pm$16.0  &\textbf{65.7$\pm$18.9}  &61.2$\pm$27.5  &51.9$\pm$23.3  &18.5$\pm$19.5   &11.8$\pm$7.15  &5.47$\pm$3.77    &16.0$\pm$15.1   &14.7$\pm$7.58     &21.6$\pm$10.1    \\
			\hline
			VarDA& \textbf{47.0$\pm$16.6} &63.1$\pm$14.1 &\textbf{73.8$\pm$14.1}  &\textbf{71.1$\pm$9.30} &\textbf{73.4$\pm$7.55}   &\textbf{4.73$\pm$1.99} &\textbf{5.33$\pm$1.60}  &\textbf{4.30$\pm$2.53} &\textbf{6.97$\pm$3.28}    &\textbf{4.56$\pm$1.66}  \\
			\hline
		\end{tabular}
	}
	
\end{table*}

\subsection{Data and experimental setup}	\label{config}

We denote the proposed VAE-based method with explicit regularization for domain adaptation as VarDA. 
We used two data sets for experiments. One is the Multi-Modality Whole Heart Segmentation (MM-WHS) Challenge dataset$\footnote{https://zmiclab.github.io/mmwhs}$ \cite{journal/mia/Zhuang2016, zhuang2019evaluation}, and the other is the Multi-Sequence Cardiac MR Segmentation (MS-CMRSeg) Challenge  dataset$\footnote{https://zmiclab.github.io/mscmrseg19}$ \cite{journal/pami/Zhuang2019}.

\textbf{MM-WHS dataset}: The organizers provided 20 MR and 20 CT 3D images with gold standard segmentation, which are not paired.
These images were collected from different patients and different clinical sites.
For evaluation, we included the following five structures for segmentation: the right atrium blood cavity (RA), the right ventricle blood cavity (RV), the left atrium blood cavity (LA), the left ventricle blood cavity (LV), and the myocardium of the left ventricle (MYO). We employed the CT images as the source domain, and the MR images as the target.
For comparison, we followed the experiment protocols in \cite{dou2018pnp,chen2019synergistic}, and randomly split the images from both domains into the training set consisting of 16 subjects and the test set of the remaining 4 subjects.
For convenience, all images were rigidly aligned and resampled into $1\times1\times1$ mm.
For 2D experiments, we used the axial slices, which were intensity normalized and cropped with an ROI of $192\times192$ pixel.

\textbf{MS-CMRSeg dataset}: The challenge provided three CMR sequences, i.e., the LGE , bSSFP and T2 images, and the target of this challenge is to segment RV, LV and MYO of the LGE CMR images. 
Hence, we used the LGE images as the target domain. 
Also, the  bSSFP CMR covers more similar part of the heart with the LGE images and contains more slices than T2, we therefore chose bSSFP images as the source domain.
The organizers provided 45 bSSFP CMR images  with 35 of them having gold standard segmentation of  RV, LV and MYO. The target data consisted of 5 annotated LGE CMR images for validation and 40 images without ground truth for test.
The bSSFP images consist of 8 to 12 contiguous slices with in-plane resolution of $1.25\times1.25$ mm, covering the full ventricles from the apex to the basal plane of the mitral valve.
The LGE CMR images are composed of 10 to 18 slices with in-plane resolution of $0.75\times0.75$ mm, covering the main body of the ventricles.
As the proposed method was for unsupervised domain adaptation problem,  we shuffled the  bSSFP CMR and LGE CMR images, which were collected from the same subjects, to make them unpaired.
For experiments, all images were intensity normalized, resized to be the same resolution of $1.0\times1.0$ mm and cropped with an ROI of $192\times192$ pixel.

The dice metric (Dice) and average symmetric surface distance (ASSD) were employed to evaluate the performance of the proposed method \cite{dou20173d}.
Dice measures the overlap between the ground truth and predicted volume or area, and
ASSD is used to assess the segmentation accuracy at the boundaries \cite{heimann2009comparison}.

\begin{figure}[h!]
	\centering
	\includegraphics[width=0.5\textwidth,height=2.5in]{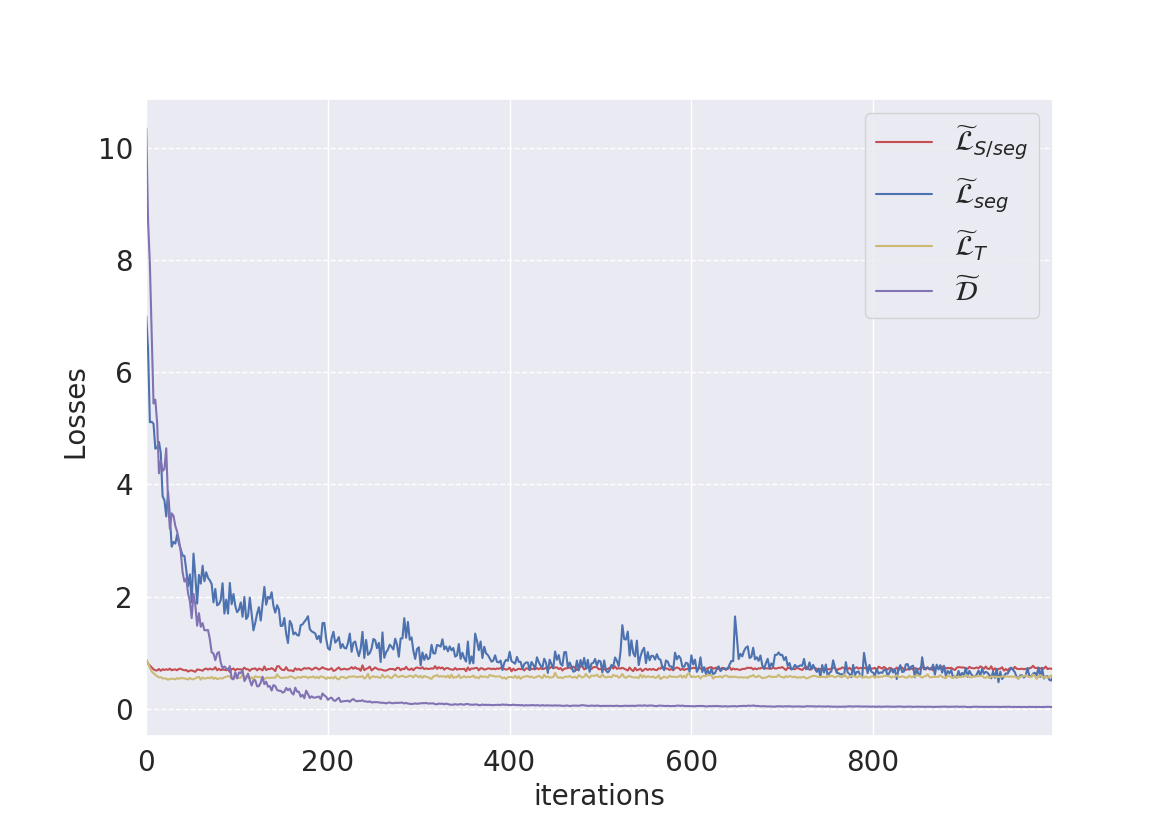}
	\caption{The losses of VarDA during training on MS-CMR dataset.
	}
	\label{fig:loss}
\end{figure}

For experiments, we designed a U-net\cite{ronneberger2015u} based network for multi-scale segmentation. The details of the network structure are presented in the Supplementary Material.
We adopted Adam optimizer \cite{article2014Kingma} for the training with batch size of 10. 
		The initial learning rate was set to be $10^{-4}$, and was decreased with a stepped decay rate of 0.9 every 150 iterations.
We empirically valued the trade-off parameters in \zxhrefeq{2.16}. To achieve this, we validated VarDA on MS-CMRSeg dataset with $\alpha_1=1$, $\alpha_2=10^{-1},10^{0},10^{1}$, $\alpha_3=10^{0},10^{-1},10^{-2},10^{-3},10^{-4},10^{-5}$ $,10^{-6}$. 
We found that the setting of $\alpha_2=1$ and $\alpha_3=10^{-2}$ was a suitable choice, and we kept this parameter setting in the following experiments.
We further plotted the loss terms of VarDA during training  in Fig.\ref{fig:loss}. As the segmentation loss from source domain contributed largely to the model optimization, we separated the loss of the source VAE, i.e., $\widetilde{\mathcal{L}}_S$, into two parts, including the segmentation loss (denoted by $\widetilde{\mathcal{L}}_{seg}$) and the remaining loss (denoted by $\widetilde{\mathcal{L}}_{S/seg}$). As shown in Fig.\ref{fig:loss}, the regularization term $\widetilde{\mathcal{D}}$ dropped fast and converged quickly.
Moreover, $\widetilde{\mathcal{L}}_{S/seg}$ and its counterpart in target domain, i.e., $\widetilde{\mathcal{L}}_{T}$, even converged in the first 100 iterations.

\subsection{Performance of the proposed method}
\label{comparison}

We compared the proposed VarDA with three state-of-the-art methods, i.e., PnP-AdaNet (also denoted as PnP-Ada for short) \cite{dou2018pnp}, SIFA \cite{chen2019synergistic} and CFDnet \cite{Wu2020TMI}.
We also included the method solely trained on the source data without adaptation, which is referred to as NoAdapt.

\begin{figure*}[!t]
	\centering
	\includegraphics[width=0.8\textwidth,height=3.0in]{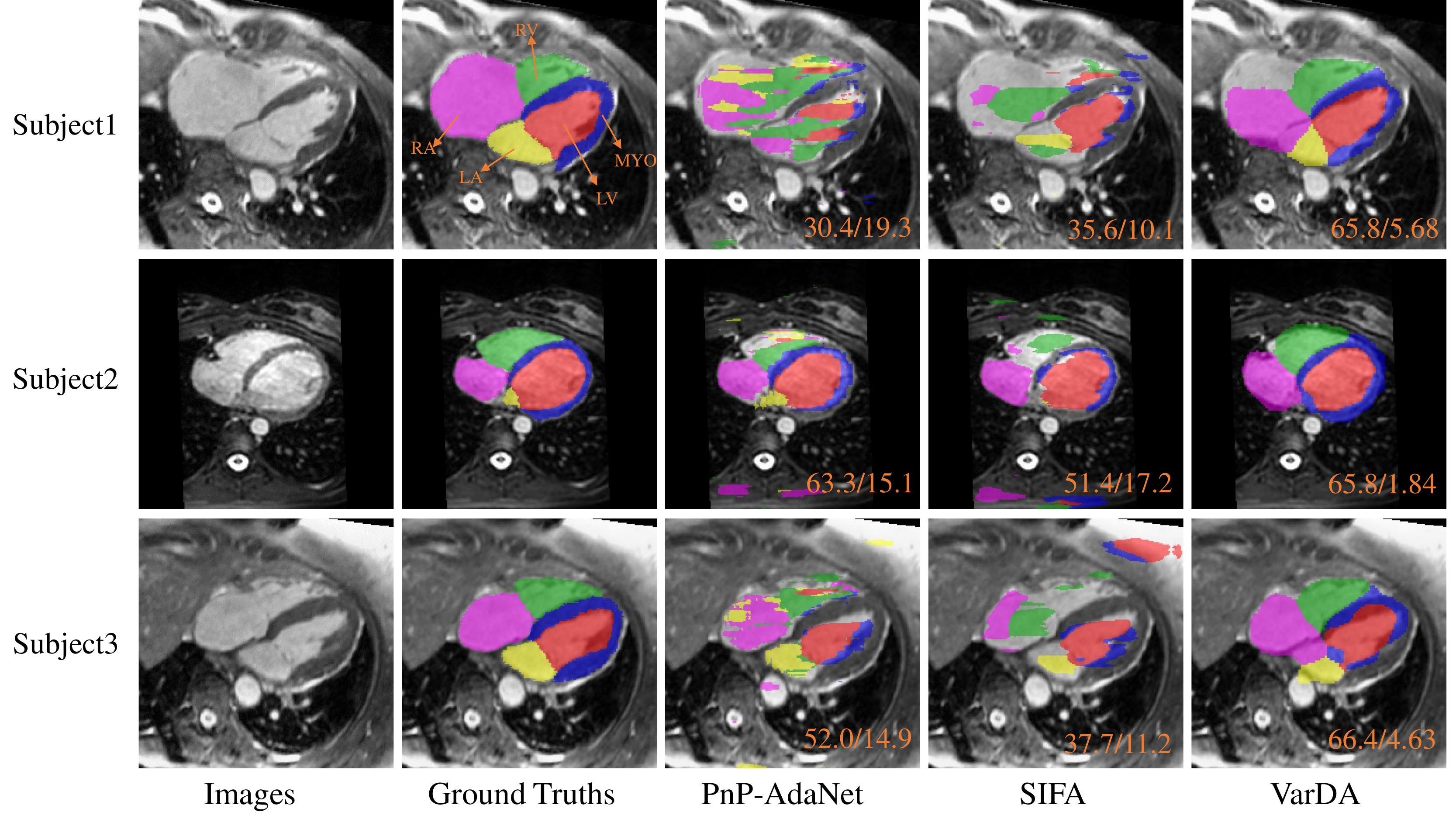}
	\caption{Visualization of 2D MR image segmentation results. Subject 1, 2, 3 are three slice cases around median Dice score from VarDA, respectively. The cardiac structure of MYO, LV, LA, RV, RA are indicated  in blue, red, yellow, green and purple color, respectively. The average Dice(\%)/ASSD(mm) value is in orange.
	}
	\label{fig:MR}
\end{figure*}

\begin{table*}[!t]
	\centering
	\begin{threeparttable}[b]
		\caption{Performance comparison of methods on LGE CMR images  from MS-CMRSeg dataset with bSSFP images as the source domain. Here, in row Best Result, we present the best Dice scores reported from the MS-CMRSeg Challenge, which did not use same setting of unsupervised domain adaptation as this work.}
		\label{bssfp to lge table}
		{\tiny
			\begin{tabular}{|l|c|c|c|c|c|c|}
				\hline
				\multirow{2}{*}{methods} & \multicolumn{3}{c|}{Dice (\%)} & \multicolumn{3}{c|}{ASSD (mm)}  \\
				\cline{2-7}
				&MYO  & LV & RV  & MYO  & LV & RV   \\
				\hline
				NoAdapt&14.50$\pm$20.18   &34.51$\pm$31.62  &31.10$\pm$26.30  &21.6$\pm$19.4  &11.3$\pm$13.1  &14.5$\pm$17.3 \\
				\hline
				PnP-AdaNet&64.64$\pm$16.41  &78.43$\pm$16.24  &72.66$\pm$19.04 &4.64$\pm$6.41  &13.8$\pm$10.3  &5.30$\pm$5.33   \\
				\hline
				CFDnet&69.1$\pm$9.69  &86.4$\pm$5.62 &76.0$\pm$10.9 &2.46$\pm$0.840 &3.07$\pm$1.66 &4.50$\pm$2.13\\
				\hline
				SIFA&70.66$\pm$9.689  &84.62$\pm$7.760  &~\textbf{83.99$\pm$6.821}  &~2.40$\pm$1.22  &~~2.68$\pm$1.14  &\textbf{2.05$\pm$1.19}   \\
				\hline
				VarDA&\textbf{73.03$\pm$8.316}  &\textbf{88.06$\pm$4.832}  &78.47$\pm$14.86   &~~\textbf{1.73$\pm$0.560}  &~\textbf{2.55$\pm$1.18}  &~3.51$\pm$2.24   \\
				\hline
				\hline
				\tabincell{c}{Best Result\tnote{1}}&82.7$\pm$6.0  &92.2$\pm$3.5  &~87.4$\pm$ 5.7 &- &- &- \\ 
				\hline
		\end{tabular}}
		\begin{tablenotes}
			\item[1] 	{\tiny Challenge result: http://www.sdspeople.fudan.edu.cn/zhuangxiahai/0/mscmrseg19/result.html}
		\end{tablenotes}
	\end{threeparttable}
\end{table*}

\subsubsection{MM-WHS results}

Table \ref{ct to mr table} reports the segmentation results of 3D MR images, where the CT images were used as the source domain.
The segmentation was done slice-by-slice in a 2D manner.
One can see the results from NoAdapt are poor, indicating there is a large domain shift between the CT and MR images.
Both PnP-AdaNet and VarDA achieve domain adaptation in a latent feature space, but the latter obtained evidently better accuracies.
SIFA, a method that conducts domain adaptation from both the image and feature perspectives, nevertheless performed worse compared to the proposed VarDA. Particularly, on the right atrium and right ventricle, SIFA had over 20\% Dice score deficit.
SIFA largely depends on the quality of the fake target image translated from the source data  \cite{chen2019synergistic}.
Since the  textures of MR images are more complex than that of the CT images, translating CT  to MR  with the same semantic contents is more difficult than the other way around.
Hence, in their work \cite{chen2019synergistic} SIFA obtained much better accuracies on CT images when MR images were used as the source domain.
For CFDnet, as it failed in the segmentation of all structures, we did not report the results in the table. This indicates that the CF distance could be challenged in domain adaptation, when the structures to be segmented are over complex.
In addition, the average Dice score of the eleven supervised methods reported in \cite{zhuang2019evaluation} ranged in $(0.674, 0.874)$ with mean value as 0.824. We provide this result here solely for reference.


Additionally, to compare with the results reported in \cite{dou2018pnp} and \cite{chen2019synergistic}, which are all in 2D,
we evaluated the performance of these methods using the 2D images extracted from the axial view of the 3D images around the center of the heart.
The results are presented in Table \ref{ct to mr slice table}.
Again, the proposed VarDA performed better in general.
VarDA achieved better Dice scores and ASSD values than PnP-AdaNet with statistical significance ($p<0.01$) for all five structures. When compared with SIFA, VarDA performed better on MYO, LV, RA and RV ($p<0.01$) in both Dice scores and ASSD values. For LA, they achieved similar Dice ($p=0.1563$) and ASSD ($p=0.7558$).

Fig. \ref{fig:MR}  visualizes the segmentation results from 3 cases, which were the median cases of VarDA according to their average Dice scores. 
One can see that the domain adaptation methods are generally challenged in this task, though the proposed method demonstrated more semantically meaningful results.
As shown in Table \ref{ct to mr slice table}, NoAdapt totally failed in this segmentation task, and thus the figures can not represent the accuracies in this case. Indeed RV could be more difficult to segment than other structures when using domain adaptation techniques. The reason could be that the contrast between RV and RA was low, while others were higher. Hence, the process of domain adaptation might function better on the structures of LV and LA, while confuse RV and RA. As shown in Fig.\ref{fig:MR}, RV was commonly predicted to be in the region of RA by PnP-AdaNet and SIFA. Note that previous works \cite{zhuang2019evaluation,journal/pami/Zhuang2019}, and challenges (such as MM-WHS and MS-CMRSeg) have shown that RV was difficult to segment.

\begin{figure*}[!t]
	\centering
	\includegraphics[width=0.8\textwidth]{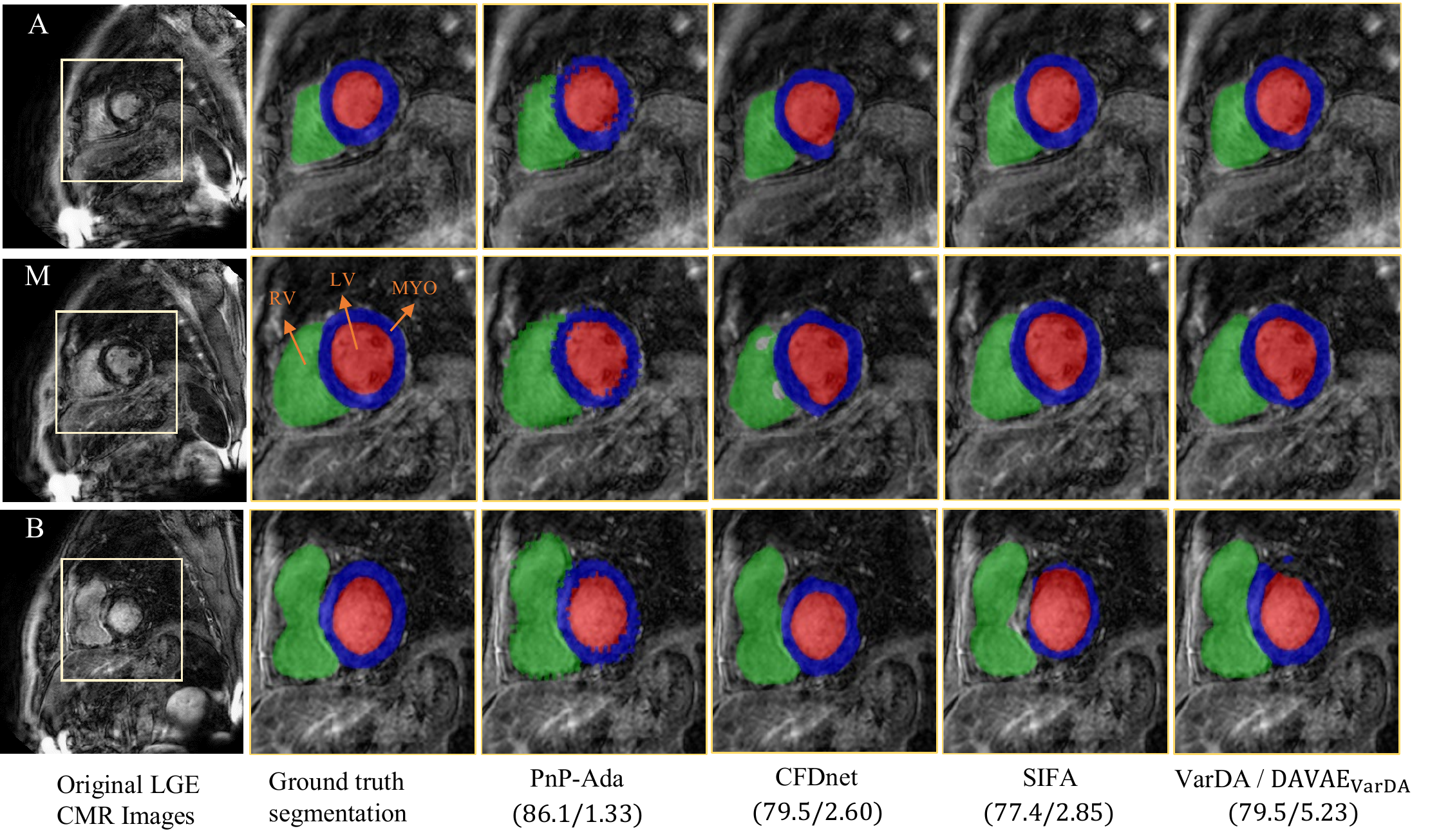}
	\caption{
		Visualization of 2D LGE MR cardiac image segmentation results. These cases are the apical (indicated with \textbf{A} in the column of Images), middle (\textbf{M}) and basal (\textbf{B}) slices from the subject with median Dice score from VarDA.
		The cardiac structure of MYO, LV, RV are indicated  in blue, red and green color respectively. 
		Note that the sub-figures of segmentation results are the zoom-in and cropped images for better visualization.
		The average Dice(\%)/ASSD(mm) values of this case from the three methods are in brackets.
		Note that VarDA is the same as DAVAE$_{\mathrm{VarDA}}$.
	}
	\label{fig:LGE}
\end{figure*}

\begin{table*}
	\centering
	\caption{Performance of the proposed framework with different inputs or numbers of convolution layers in the decoder. The methods were tested on LGE CMR segmentation, where the bSSFP CMR was used as the source domain.}
	\label{bssfp to lge recon table}
	\resizebox{350pt}{36pt}{\begin{tabular}{|l|c|c|c|c|c|c|}
			\hline
			\multirow{2}{*}{No. of layers } & \multicolumn{3}{c|}{Dice (\%)} & \multicolumn{3}{c|}{ASSD (mm)}  \\
			\cline{2-7}
			&MYO  & LV & RV  & MYO  & LV & RV   \\
			\hline
			VarDA ($N$=0)&70.70$\pm$9.503  &86.62$\pm$5.762  &77.87$\pm$10.29   &1.92$\pm$0.682  &2.63$\pm$1.23  &4.81$\pm$2.61   \\
			\hline
			VarDA ($N$=3)&70.73$\pm$9.453 &86.61$\pm$6.271  &77.38$\pm$11.26    &1.86$\pm$0.661 &2.60$\pm$1.21 & 4.21$\pm$2.01  \\
			\hline
			VarDA ($N$=7)&\textbf{73.03$\pm$8.316}  &\textbf{88.06$\pm$4.832} &78.47$\pm$14.86   &\textbf{1.73$\pm$0.556}  &\textbf{2.55$\pm$1.18 } &3.51$\pm$2.24   \\
			\hline
			VarDA ($N$=11)&72.86$\pm$9.101  &87.26$\pm$6.071  &78.27$\pm$10.08    &1.89$\pm$0.854  &2.72$\pm$1.68  &3.90$\pm$2.12  \\
			\hline
			VarDA$_{\mathrm{weak}}$ ($N=3$)&70.15$\pm$9.409  &86.55$\pm$5.212  &74.93$\pm$12.03   &2.04$\pm$0.722  &3.10$\pm$1.16  &5.12$\pm$2.32   \\
			\hline
			VarDA$_{\mathrm{weak}}$ ($N=7$)&71.80$\pm$8.492  &86.75$\pm$5.571  &78.11$\pm$10.75   &2.00$\pm$0.839  &2.93$\pm$1.54  &3.80$\pm$1.91   \\
			\hline
			VarDA$_{\mathrm{weak}}$ ($N=11$)&72.16$\pm$9.825  &87.18$\pm$5.854  &\textbf{79.08$\pm$10.75}   &1.99$\pm$0.908  &2.71$\pm$1.40  &\textbf{3.43$\pm$2.04}   \\
			\hline
	\end{tabular}}
\end{table*}
\begin{figure*}[!t]
	\centering
	\includegraphics[width=1 \textwidth]{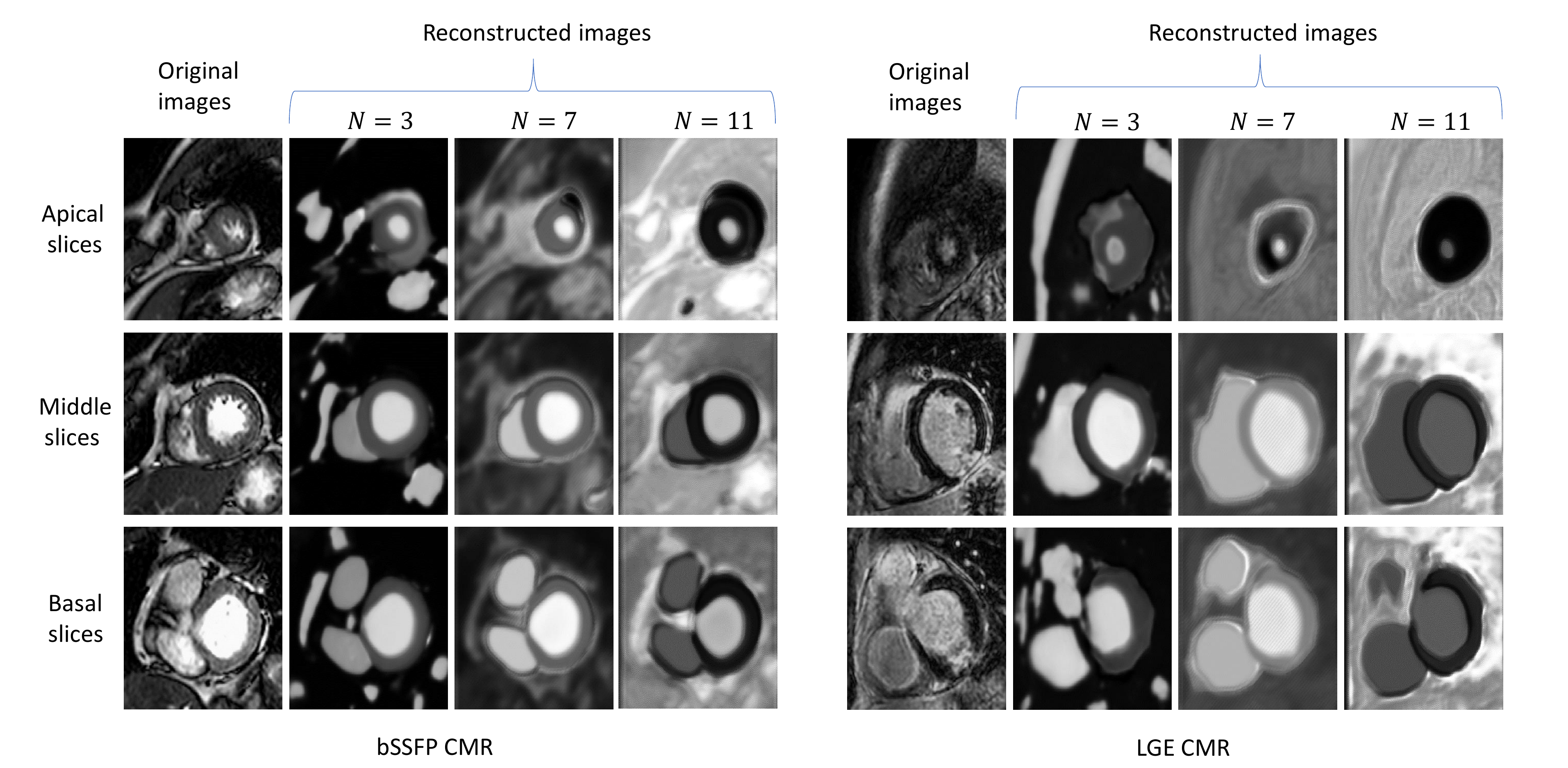}\\[-1ex]
	\caption{Typical examples of reconstructed images from the encoder with different numbers of layers in the proposed framework.}
	\label{fig:ReconImg}
\end{figure*}

\subsubsection{MS-CMRSeg results}
Table \ref{bssfp to lge table} presents the Dice and ASSD of LGE CMR segmentation, where the bSSFP images were used as the source domain.
The segmentation was done slice-by-slice in a 2D manner.
Fig. \ref{fig:LGE} visualizes the segmentation results of a subject, which was the median case of VarDA according to the average Dice score.

Compared with the task for domain adaptation between cardiac CT and MR images, NoAdapt in this task performed rather well, indicating that the domain shift between bSSFP and LGE images is much smaller than that between CT and MR. Therefore, the domain adaptation between bSSFP and LGE is easier.
Compared to PnP-AdaNet, VarDA obtained significantly better Dice scores and ASSD values on all structures ($p<0.01$).
Compared to SIFA, VarDA achieved better accuracies on MYO and LV segmentation but worse on RV.
Compared to CFDnet, VarDA obtained comparable average Dice score ($p= 0.05042$), and significantly better average ASSD value ($p=0.001487$).
	In addition, for reference we provide the results of MS-CMRSeg challenge here. The Dice scores of the top ten methods on the leaderboard, which tackled paired data, ranged in $(0.854, 0.922)$ for LV, $(0.713, 0.827)$ for Myo and $(0.792, 0.874)$ for RV.
	We further provided the best Dice scores reported from the MS-CMRSeg Challenge in Table \ref{bssfp to lge table} for reference.

\subsection{Effect of reconstruction}
\label{recon}

This study investigates the role and effect of the reconstruction model to the segmentation
in the proposed domain adaptation framework.
It is arguable that the reconstruction mechanism of the decoder helps constrain the anatomic shape,
which leads to a better performance of the segmentor.
To verify this, we first altered the modeling capability of the reconstruction model, i.e., the number of convolution layers in the decoder, denoted as $N$, and compared the segmentation performance of the framework.
Four schemes were evaluated, i.e., with $N$=0, 3, 7 and 11, where $N=0$ means the reconstruction model is removed from the proposed framework.
Then, we further studied the proposed framework with a reconstruction model that solely takes the latent features as input without the segmentation, namely the decoder in \zxhrefeq{2.8} becomes $E_{q_{\phi_S}(z|x)}[ p_{\theta_S}(x|z)]$.
This is a weaker version of the proposed method and thus is referred to as VarDA$_{\mathrm{weak}}$.
We conducted this study on the LGE segmentation using bSSFP CMR as the source domain.

Table \ref{bssfp to lge recon table} provides the Dice and ASSD results of LGE CMR segmentation.
VarDA ($N=3$) with three layers for the decoder network did not perform evidently better than VarDA ($N=0$) ($p=0.6836$ for Dice  and $p=0.0236$ for  ASSD), indicating the gain from the small reconstruction model was limited.
However, when seven layers were used, where the decoder had a much greater reconstruction capacity,  VarDA ($N=7$) achieved much better segmentation accuracies ($p<0.01$ for both Dice and ASSD).
When the number of layers increased to eleven, VarDA ($N=11$) achieved similar results with VarDA ($N=7$) ($p=0.5370$ for Dice  and $p=0.0518$ for  ASSD), probably due to the fact that the reconstruction capacity had reached its upper limit, and no gain could be pursued.
For VarDA$_{\mathrm{weak}}$, similar results were observed.
We have provided visualization examples of the reconstruction results in the Supplementary Material, which agreed with the quantitative results in Table \ref{bssfp to lge recon table}.

With the features being updated to be domain-invariant during the training stage, the generated target label would be more accurate than those of NoAdapt.  Moreover,  with the optimization process to minimize the reconstruction error, the anatomic structure of the pseudo labels can be further improved.
\Zxhreftb{bssfp to lge recon table} presents the results of the weak form of VarDA with seven-layer CNN reconstructor, i.e., VarDA$_{\mathrm{weak}}$ ($N=7$), where the reconstructor does not have the input of the segmentation.
One can see that VarDA ($N=7$) obtained evidently better Dice scores ($p<0.01$) and ASSD values ($p=0.011$) than VarDA$_{\mathrm{weak}}$ ($N=7$).
This confirms that VarDA, with reconstruction network having inputs of both the latent features and  (pseudo) labels, could achieve better segmentation performance, thanks to the constraints and knowledge of anatomic shapes in the segmentation.

Fig. \ref{fig:ReconImg} visualizes the reconstruction results from 3 bSSFP CMR images and 3 LGE CMR  images.
		One can see that the reconstructed images from the decoders with higher capacity networks, i.e.,  $N=7$ and $N=11$, contain more meaningful textures, when compared with the images reconstructed from VarDA ($N=3$).
		However, the texture details reconstructed from VarDA ($N=7$) and VarDA ($N=11$) are comparable, which agrees with the quantitative results of segmentation accuracies in Table V of the original manuscript.

\subsection{Comparisons with adversarial training}
\label{adversarial ablation}

In this section, we compared the proposed method with two other approaches using adversarial training .
Here, the adversarial training with joint distributions is referred to as DAVAE$_{\mathrm{AdvJoint}}$ , where we concatenate the latent features and the prediction label for discrimination.
Similarly, the adversarial training without joint distributions is referred to as DAVAE$_{\mathrm{Adv}}$, where we only use the latent features for discrimination.
For comparison, we also denote the proposed method as DAVAE$_{\mathrm{VarDA}}$.
We applied these three methods to the segmentation of LGE CMR where bSSFP CMR was used as the source domain.
The results shown that the proposed method performs much better than the adversarial approaches in the same VAE-based U-Net framework.
As shown in the work of Kamnitsas et al. \cite{kamnitsas2017unsupervise} and Dou et al. \cite{dou2018pnp}, deep features of multi-layers were much more effective for domain adaptation than shallow features.
		While in the VAE-based framework, the features for adaptation were shallower than those of PnP-AdaNet.
		Moreover, the skip connections in the U-Net of the framework concatenated the shallow features with deep ones, which further lead the features ineffective for adversarial training based methods.
Details of the framework for adversarial training and the comparison results are provided in the Supplementary Material.

\begin{figure}[h!]
	\centering
	\includegraphics[width=0.5\textwidth,height=2.5in]{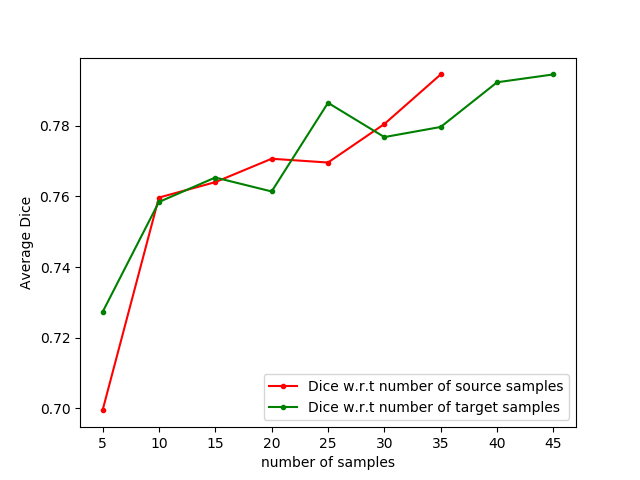}
	\caption{Model performance with regard to the number of samples during training. The red line shows the results with all the 45 LGE CMR images (target domain) and different number of bSSFP CMR images (source domain). The green line shows the  results with all the 35 bSSFP CMR images and different number of LGE CMR images.
	}
	\label{fig:dice_num}
\end{figure}

\begin{figure}[h!]
	\centering
	\subfloat[Before domain adaptation.]{\includegraphics[width=1.8in]{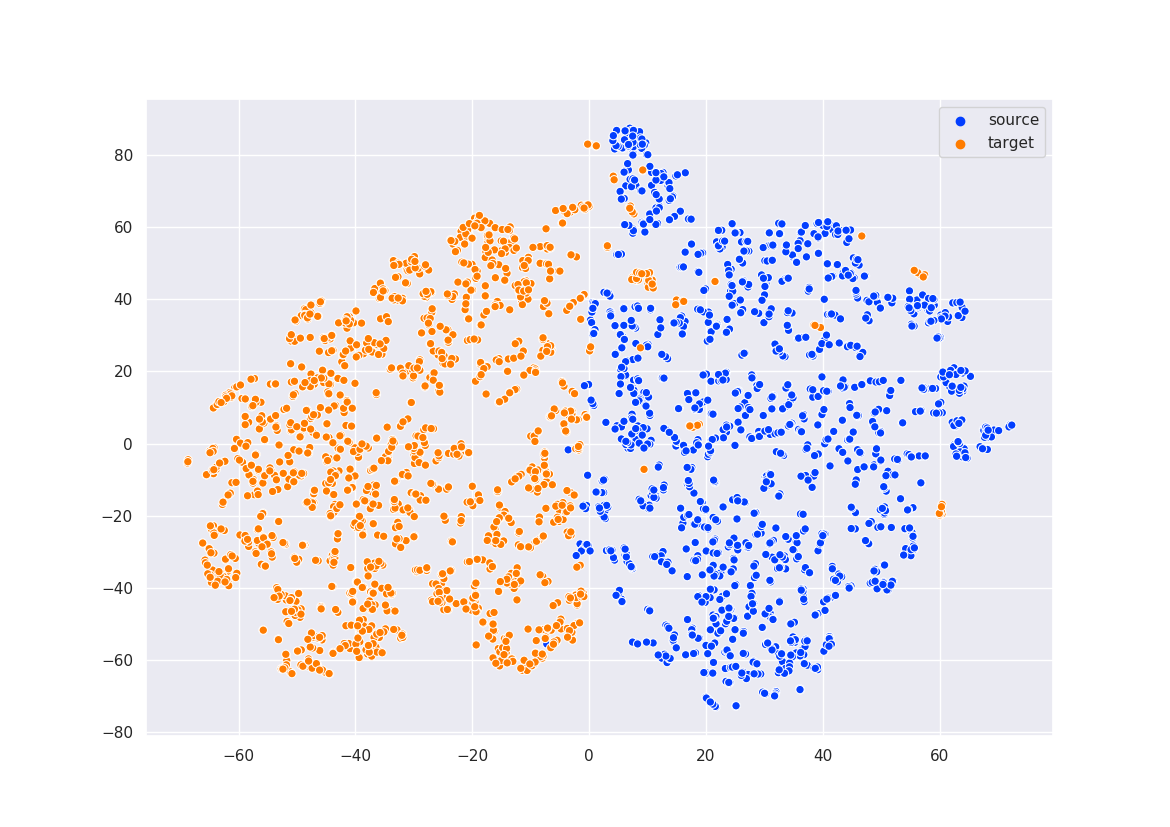}
		\label{fig:tsne1}}
	\subfloat[After domain adaptation.]{\includegraphics[width=1.8in]{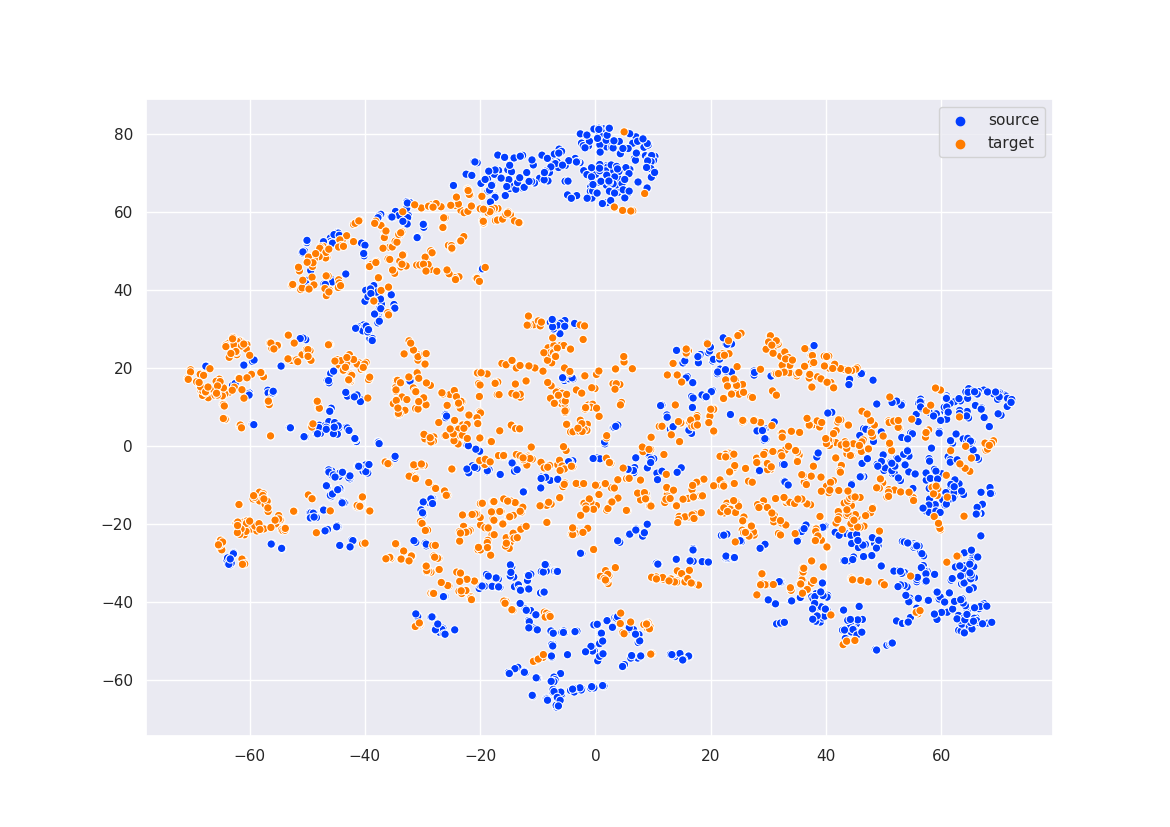}
		\label{fig:tsne2}}
	\caption{T-SNE plot of features for bSSFP (blue) $\rightarrow$ LGE CMR (orange) images via VarDA.}
	\label{fig:tsne}
\end{figure}

\subsection{Performance with different sample complexity}
To validate the model performance with different sample complexity, we conducted two set of ablation experiments. We first used all the 45 LGE CMR images  as the target data, and set the number of source data, i.e., the bSSFP CMR images, to be 5, 10, 15, 20, 25, 30 and 35, respectively. As Fig. \ref{fig:dice_num} showed, the performance of the model improved and even did not show convergence with the increasing number of the source samples.
We further did the experiments with all the source images, and set the number of target data to be 5, 10, 15, 20, 25, 30, 35, 40 and 45, respectively. As shown in Fig. \ref{fig:dice_num}, the performance of the model also improved with the increasing number of the target samples, though not as stably as the improvement with different source data.
The results demonstrated that both the diversities of source and target data were important for the model training.

To see how the distribution discrepancy was reduced by VarDA, we further visualized the features by T-SNE \cite{JMLRv9vandermaaten08a}.
As shown in Fig.\ref{fig:tsne}, the data points in two domains showed strong clustering before adaptation.
After training with VarDA, the representations learned by the model became more indistinguishable between the source and target domains.

\section{Conclusion}
\label{conclude}

In this work, we proposed a new VAE-based framework which drives two domains to one parameterized distribution, and introduced a regularization for the variational approximation in an explicit formulation.
		Minimization of this regularization has been shown to be efficient in correcting domain shifts in the segmentation tasks for unsupervised domain adaptation.
We validated the proposals using two segmentation tasks, i.e., the cross-modality (CT and MR) whole heart segmentation and the cross-sequence (bSSFP and LGE) CMR segmentation.
The results showed that the explicit regularization was more effective, compared to the adversarial training, and the proposed domain adaptation method achieved state-of-the-art performance in the two segmentation tasks.





\bibliographystyle{IEEEtran}
\bibliography{IEEEabrv,TMI2021}

\begin{thebibliography}{10}
\providecommand{\url}[1]{#1}
\csname url@samestyle\endcsname
\providecommand{\newblock}{\relax}
\providecommand{\bibinfo}[2]{#2}
\providecommand{\BIBentrySTDinterwordspacing}{\spaceskip=0pt\relax}
\providecommand{\BIBentryALTinterwordstretchfactor}{4}
\providecommand{\BIBentryALTinterwordspacing}{\spaceskip=\fontdimen2\font plus
\BIBentryALTinterwordstretchfactor\fontdimen3\font minus
  \fontdimen4\font\relax}
\providecommand{\BIBforeignlanguage}[2]{{%
\expandafter\ifx\csname l@#1\endcsname\relax
\typeout{** WARNING: IEEEtran.bst: No hyphenation pattern has been}%
\typeout{** loaded for the language `#1'. Using the pattern for}%
\typeout{** the default language instead.}%
\else
\language=\csname l@#1\endcsname
\fi
#2}}
\providecommand{\BIBdecl}{\relax}
\BIBdecl

\bibitem{KARAMITSOS20091407}
T.~D. Karamitsos, J.~M. Francis, S.~Myerson, J.~B. Selvanayagam, and
  S.~Neubauer, ``The role of cardiovascular magnetic resonance imaging in heart
  failure,'' \emph{Journal of the American College of Cardiology}, vol.~54,
  no.~15, pp. 1407 -- 1424, 2009.

\bibitem{petitjean2011review}
C.~Petitjean and J.-N. Dacher, ``A review of segmentation methods in short axis
  cardiac {MR} images,'' \emph{Medical image analysis}, vol.~15, no.~2, pp.
  169--184, 2011.

\bibitem{shimodaira2000improving}
H.~Shimodaira, ``Improving predictive inference under covariate shift by
  weighting the log-likelihood function,'' \emph{Journal of statistical
  planning and inference}, vol.~90, no.~2, pp. 227--244, 2000.

\bibitem{babenko2014neural}
A.~Babenko, A.~Slesarev, A.~Chigorin, and V.~Lempitsky, ``Neural codes for
  image retrieval,'' in \emph{Computer Vision -- ECCV 2014}.\hskip 1em plus
  0.5em minus 0.4em\relax Cham: Springer International Publishing, 2014, pp.
  584--599.

\bibitem{chu2016best}
B.~Chu, V.~Madhavan, O.~Beijbom, J.~Hoffman, and T.~Darrell, ``Best practices
  for fine-tuning visual classifiers to new domains,'' in \emph{Computer Vision
  -- ECCV 2016 Workshops}.\hskip 1em plus 0.5em minus 0.4em\relax Cham:
  Springer International Publishing, 2016, pp. 435--442.

\bibitem{oquab2014learning}
M.~Oquab, L.~Bottou, I.~Laptev, and J.~Sivic, ``Learning and transferring
  mid-level image representations using convolutional neural networks,'' in
  \emph{Proceedings of the IEEE conference on computer vision and pattern
  recognition}, 2014, pp. 1717--1724.

\bibitem{raina2007self}
R.~Raina, A.~Battle, H.~Lee, B.~Packer, and A.~Y. Ng, ``Self-taught learning:
  transfer learning from unlabeled data,'' in \emph{Proceedings of the 24th
  international conference on Machine learning}.\hskip 1em plus 0.5em minus
  0.4em\relax ACM, 2007, pp. 759--766.

\bibitem{thrun2012learning}
S.~Thrun and L.~Pratt, \emph{Learning to learn}.\hskip 1em plus 0.5em minus
  0.4em\relax Springer Science \& Business Media, 2012.

\bibitem{Csurka2017}
G.~Csurka, \emph{A Comprehensive Survey on Domain Adaptation for Visual
  Applications}.\hskip 1em plus 0.5em minus 0.4em\relax Cham: Springer
  International Publishing, 2017, pp. 1--35.

\bibitem{dou2018pnp}
Q.~{Dou}, C.~{Ouyang}, C.~{Chen}, H.~{Chen}, B.~{Glocker}, X.~{Zhuang}, and
  P.~{Heng}, ``Pn{P}-{A}da{N}et: Plug-and-play adversarial domain adaptation
  network at unpaired cross-modality cardiac segmentation,'' \emph{IEEE
  Access}, vol.~7, pp. 99\,065--99\,076, 2019.

\bibitem{chen2019synergistic}
C.~{Chen}, Q.~{Dou}, H.~{Chen}, J.~{Qin}, and P.~A. {Heng}, ``Unsupervised
  bidirectional cross-modality adaptation via deeply synergistic image and
  feature alignment for medical image segmentation,'' \emph{IEEE Transactions
  on Medical Imaging}, vol.~39, no.~7, pp. 2494--2505, 2020.

\bibitem{fernando2013unsupervised}
B.~Fernando, A.~Habrard, M.~Sebban, and T.~Tuytelaars, ``Unsupervised visual
  domain adaptation using subspace alignment,'' in \emph{Proceedings of the
  2013 IEEE International Conference on Computer Vision}, ser. ICCV '13.\hskip
  1em plus 0.5em minus 0.4em\relax Washington, DC, USA: IEEE Computer Society,
  2013, pp. 2960--2967.

\bibitem{sun2016return}
B.~Sun, J.~Feng, and K.~Saenko, ``Return of frustratingly easy domain
  adaptation,'' in \emph{Proceedings of the Thirtieth AAAI Conference on
  Artificial Intelligence}, ser. AAAI'16.\hskip 1em plus 0.5em minus
  0.4em\relax AAAI Press, 2016, pp. 2058--2065.

\bibitem{tommasi2013frustratingly}
T.~Tommasi and B.~Caputo, ``Frustratingly easy {NBNN} domain adaptation,'' in
  \emph{Proceedings of the 2013 IEEE International Conference on Computer
  Vision}, ser. ICCV '13.\hskip 1em plus 0.5em minus 0.4em\relax Washington,
  DC, USA: IEEE Computer Society, 2013, pp. 897--904.

\bibitem{arjovsky2017wasserstein}
M.~Arjovsky, S.~Chintala, and L.~Bottou, ``Wasserstein {GAN},'' \emph{CoRR},
  vol. abs/1701.07875, 2017.

\bibitem{ganin2016domain}
Y.~Ganin, E.~Ustinova, H.~Ajakan, P.~Germain, H.~Larochelle, F.~Laviolette,
  M.~Marchand, and V.~Lempitsky, ``Domain-adversarial training of neural
  networks,'' \emph{The Journal of Machine Learning Research}, vol.~17, no.~1,
  pp. 2096--2030, 2016.

\bibitem{goodfellow2014generative}
I.~Goodfellow, J.~Pouget-Abadie, M.~Mirza, B.~Xu, D.~Warde-Farley, S.~Ozair,
  A.~Courville, and Y.~Bengio, ``Generative adversarial nets,'' in
  \emph{Advances in neural information processing systems}, 2014, pp.
  2672--2680.

\bibitem{kamnitsas2017unsupervised}
K.~Kamnitsas, C.~Baumgartner, C.~Ledig, V.~Newcombe, J.~Simpson, A.~Kane,
  D.~Menon, A.~Nori, A.~Criminisi, D.~Rueckert \emph{et~al.}, ``Unsupervised
  domain adaptation in brain lesion segmentation with adversarial networks,''
  in \emph{International conference on information processing in medical
  imaging}.\hskip 1em plus 0.5em minus 0.4em\relax Springer, 2017, pp.
  597--609.

\bibitem{heusel2017gans}
M.~Heusel, H.~Ramsauer, T.~Unterthiner, B.~Nessler, and S.~Hochreiter, ``{GANs}
  trained by a two time-scale update rule converge to a local nash
  equilibrium,'' in \emph{Advances in Neural Information Processing Systems
  30}.\hskip 1em plus 0.5em minus 0.4em\relax Curran Associates, Inc., 2017,
  pp. 6626--6637.

\bibitem{long2015learning}
M.~Long, Y.~Cao, J.~Wang, and M.~Jordan, ``Learning transferable features with
  deep adaptation networks,'' in \emph{Proceedings of the 32nd International
  Conference on Machine Learning}, ser. Proceedings of Machine Learning
  Research, vol.~37.\hskip 1em plus 0.5em minus 0.4em\relax Lille, France:
  PMLR, 07--09 Jul 2015, pp. 97--105.

\bibitem{long2017deep}
M.~Long, H.~Zhu, J.~Wang, and M.~I. Jordan, ``Deep transfer learning with joint
  adaptation networks,'' in \emph{Proceedings of the 34th International
  Conference on Machine Learning}, ser. Proceedings of Machine Learning
  Research, vol.~70.\hskip 1em plus 0.5em minus 0.4em\relax International
  Convention Centre, Sydney, Australia: PMLR, 06--11 Aug 2017, pp. 2208--2217.

\bibitem{sun2016deep}
B.~Sun and K.~Saenko, ``Deep coral: Correlation alignment for deep domain
  adaptation,'' in \emph{European Conference on Computer Vision}.\hskip 1em
  plus 0.5em minus 0.4em\relax Springer, 2016, pp. 443--450.

\bibitem{tzeng2014deep}
E.~Tzeng, J.~Hoffman, N.~Zhang, K.~Saenko, and T.~Darrell, ``Deep domain
  confusion: Maximizing for domain invariance,'' \emph{arXiv preprint
  arXiv:1412.3474}, 2014.

\bibitem{Wu2020TMI}
F.~{Wu} and X.~{Zhuang}, ``{CF} distance: A new domain discrepancy metric and
  application to explicit domain adaptation for cross-modality cardiac image
  segmentation,'' \emph{IEEE Transactions on Medical Imaging}, pp. 1--12, 2020.

\bibitem{chartsias2017multimodal}
A.~Chartsias, T.~Joyce, M.~V. Giuffrida, and S.~A. Tsaftaris, ``Multimodal {MR}
  synthesis via modality-invariant latent representation,'' \emph{IEEE
  transactions on medical imaging}, vol.~37, no.~3, pp. 803--814, 2017.

\bibitem{van2019learning}
G.~{van Tulder} and M.~{de Bruijne}, ``Learning cross-modality representations
  from multi-modal images,'' \emph{IEEE Transactions on Medical Imaging},
  vol.~38, no.~2, pp. 638--648, Feb 2019.

\bibitem{conf/nips/ShiNPT19}
Y.~Shi, S.~Narayanaswamy, B.~Paige, and P.~H.~S. Torr, ``Variational
  mixture-of-experts autoencoders for multi-modal deep generative models,'' in
  \emph{Advances in Neural Information Processing Systems 32: Annual Conference
  on Neural Information Processing Systems 2019, NeurIPS 2019, December 8-14,
  2019, Vancouver, BC, Canada}, 2019, pp. 15\,692--15\,703.

\bibitem{conf/icml/AntelmiARL19}
L.~Antelmi, N.~Ayache, P.~Robert, and M.~Lorenzi, ``Sparse multi-channel
  variational autoencoder for the joint analysis of heterogeneous data,'' in
  \emph{Proceedings of the 36th International Conference on Machine Learning,
  {ICML} 2019, 9-15 June 2019, Long Beach, California, {USA}}, 2019, pp.
  302--311.

\bibitem{wang2017deep}
Y.~Wang, W.~Li, D.~Dai, and L.~Van~Gool, ``Deep domain adaptation by geodesic
  distance minimization,'' in \emph{Proceedings of the IEEE International
  Conference on Computer Vision}, 2017, pp. 2651--2657.

\bibitem{Debasmit2018Graph}
D.~Das and C.~S.~G. Lee, ``Graph matching and pseudo-label guided deep
  unsupervised domain adaptation,'' in \emph{Artificial Neural Networks and
  Machine Learning -- ICANN 2018}, 2018, pp. 342--352.

\bibitem{conf/aaai/YangY19}
B.~Yang and P.~C. Yuen, ``Cross-domain visual representations via unsupervised
  graph alignment,'' in \emph{The Thirty-Third {AAAI} Conference on Artificial
  Intelligence, {AAAI} 2019, The Thirty-First Innovative Applications of
  Artificial Intelligence Conference, {IAAI} 2019, The Ninth {AAAI} Symposium
  on Educational Advances in Artificial Intelligence, {EAAI} 2019, Honolulu,
  Hawaii, USA, January 27 - February 1, 2019}, 2019, pp. 5613--5620.

\bibitem{9055059Pilanci}
M.~{Pilanci} and E.~{Vural}, ``Domain adaptation on graphs by learning aligned
  graph bases,'' \emph{IEEE Transactions on Knowledge and Data Engineering},
  pp. 1--1, 2020.

\bibitem{zhang2018translating}
Z.~Zhang, L.~Yang, and Y.~Zheng, ``Translating and segmenting multimodal
  medical volumes with cycle- and shape-consistency generative adversarial
  network,'' in \emph{Proceedings of the IEEE Conference on Computer Vision and
  Pattern Recognition}, 2018, pp. 9242--9251.

\bibitem{huang2018multimodal}
X.~Huang, M.-Y. Liu, S.~Belongie, and J.~Kautz, ``Multimodal unsupervised
  image-to-image translation,'' in \emph{Proceedings of the European Conference
  on Computer Vision (ECCV)}, 2018, pp. 172--189.

\bibitem{ouyang2019data}
C.~Ouyang, K.~Kamnitsas, C.~Biffi, J.~Duan, and D.~Rueckert, ``Data efficient
  unsupervised domain adaptation for cross-modality image segmentation,'' in
  \emph{Medical Image Computing and Computer Assisted Intervention - {MICCAI}
  2019 - 22nd International Conference, Shenzhen, China, October 13-17, 2019,
  Proceedings, Part {II}}, 2019, pp. 669--677.

\bibitem{kingma2013auto}
D.~P. Kingma and M.~Welling, ``Auto-encoding variational bayes,'' \emph{CoRR},
  vol. abs/1312.6114, 2013.

\bibitem{kingma2014semi}
D.~P. Kingma, D.~J. Rezende, S.~Mohamed, and M.~Welling, ``Semi-supervised
  learning with deep generative models,'' in \emph{Proceedings of the 27th
  International Conference on Neural Information Processing Systems}, ser.
  NIPS'14, vol.~2.\hskip 1em plus 0.5em minus 0.4em\relax Cambridge, MA, USA:
  MIT Press, 2014, pp. 3581--3589.

\bibitem{sohn2015learning}
K.~Sohn, H.~Lee, and X.~Yan, ``Learning structured output representation using
  deep conditional generative models,'' in \emph{Advances in Neural Information
  Processing Systems 28}.\hskip 1em plus 0.5em minus 0.4em\relax Curran
  Associates, Inc., 2015, pp. 3483--3491.

\bibitem{walker2016uncertain}
J.~Walker, C.~Doersch, A.~Gupta, and M.~Hebert, ``An uncertain future:
  Forecasting from static images using variational autoencoders,'' in
  \emph{European Conference on Computer Vision}.\hskip 1em plus 0.5em minus
  0.4em\relax Springer, 2016, pp. 835--851.

\bibitem{grandvalet2005semi}
Y.~Grandvalet and Y.~Bengio, ``Semi-supervised learning by entropy
  minimization,'' in \emph{Advances in neural information processing systems},
  2005, pp. 529--536.

\bibitem{shu2018a}
R.~Shu, H.~H. Bui, H.~Narui, and S.~Ermon, ``A {DIRT-T} approach to
  unsupervised domain adaptation,'' in \emph{6th International Conference on
  Learning Representations, {ICLR} 2018, Conference Track Proceedings}, 2018.

\bibitem{rezende2014stochastic}
D.~J. Rezende, S.~Mohamed, and D.~Wierstra, ``Stochastic backpropagation and
  approximate inference in deep generative models,'' in \emph{Proceedings of
  the 31st International Conference on Machine Learning}, ser. Proceedings of
  Machine Learning Research, vol.~32, no.~2.\hskip 1em plus 0.5em minus
  0.4em\relax Bejing, China: PMLR, 22--24 Jun 2014, pp. 1278--1286.

\bibitem{journal/mia/Zhuang2016}
X.~Zhuang and J.~Shen, ``Multi-scale patch and multi-modality atlases for whole
  heart segmentation of {MRI},'' \emph{Medical image analysis}, vol.~31, pp.
  77--87, 2016.

\bibitem{zhuang2019evaluation}
X.~Zhuang, L.~Li, C.~Payer, D.~Štern, M.~Urschler, M.~P. Heinrich, J.~Oster,
  C.~Wang, Örjan Smedby, C.~Bian, X.~Yang, P.-A. Heng, A.~Mortazi, U.~Bagci,
  G.~Yang, C.~Sun, G.~Galisot, J.-Y. Ramel, T.~Brouard, Q.~Tong, W.~Si,
  X.~Liao, G.~Zeng, Z.~Shi, G.~Zheng, C.~Wang, T.~MacGillivray, D.~Newby,
  K.~Rhode, S.~Ourselin, R.~Mohiaddin, J.~Keegan, D.~Firmin, and G.~Yang,
  ``Evaluation of algorithms for multi-modality whole heart segmentation: An
  open-access grand challenge,'' \emph{Medical Image Analysis}, vol.~58, p.
  101537, 2019.

\bibitem{journal/pami/Zhuang2019}
X.~{Zhuang}, ``Multivariate mixture model for myocardial segmentation combining
  multi-source images,'' \emph{IEEE Transactions on Pattern Analysis and
  Machine Intelligence}, vol.~41, no.~12, pp. 2933--2946, Dec 2019.

\bibitem{dou20173d}
Q.~Dou, L.~Yu, H.~Chen, Y.~Jin, X.~Yang, J.~Qin, and P.-A. Heng, ``{3D} deeply
  supervised network for automated segmentation of volumetric medical images,''
  \emph{Medical image analysis}, vol.~41, pp. 40--54, 2017.

\bibitem{heimann2009comparison}
T.~Heimann, B.~Van~Ginneken, M.~A. Styner, Y.~Arzhaeva, V.~Aurich, C.~Bauer,
  A.~Beck, C.~Becker, R.~Beichel, G.~Bekes \emph{et~al.}, ``Comparison and
  evaluation of methods for liver segmentation from {CT} datasets,'' \emph{IEEE
  transactions on medical imaging}, vol.~28, no.~8, pp. 1251--1265, 2009.

\bibitem{ronneberger2015u}
O.~Ronneberger, P.~Fischer, and T.~Brox, ``U-net: Convolutional networks for
  biomedical image segmentation,'' in \emph{International Conference on Medical
  image computing and computer-assisted intervention}.\hskip 1em plus 0.5em
  minus 0.4em\relax Springer, 2015, pp. 234--241.

\bibitem{article2014Kingma}
D.~Kingma and J.~Ba, ``Adam: A method for stochastic optimization,'' in
  \emph{International Conference on Learning Representations}, 2014.

\bibitem{kamnitsas2017unsupervise}
K.~Kamnitsas, C.~Baumgartner, C.~Ledig, V.~Newcombe, J.~Simpson, A.~Kane,
  D.~Menon, A.~Nori, A.~Criminisi, D.~Rueckert \emph{et~al.}, ``Unsupervised
  domain adaptation in brain lesion segmentation with adversarial networks,''
  in \emph{International conference on information processing in medical
  imaging}.\hskip 1em plus 0.5em minus 0.4em\relax Springer, 2017, pp.
  597--609.

\bibitem{JMLRv9vandermaaten08a}
L.~van~der Maaten and G.~Hinton, ``Visualizing data using t-sne,''
  \emph{Journal of Machine Learning Research}, vol.~9, no.~86, pp. 2579--2605,
  2008.

\end{thebibliography}
%

\end{document}